\documentclass[journal]{IEEEtran}
\usepackage{cite}
\usepackage{graphicx,amssymb,lineno}
\usepackage{amsmath,amsfonts,amssymb}

\usepackage{algorithm}
\usepackage{algorithmic}
\usepackage[usenames]{color}
\usepackage{float}
\usepackage{mathtools}

\usepackage{graphicx,graphics,color,epsfig,subfigure,graphpap,rotate}
\usepackage{times, verbatim, subfigure, epsfig, graphicx, latexsym, amsmath}
\usepackage{url}
\usepackage{subfigure}
\usepackage{CJK}

\begin{document}

\title{Resource Allocation for Device-to-Device Communications in Multi-Cell Multi-Band Heterogeneous Cellular Networks}

\author{Yali~Chen,
        Bo~Ai,~\IEEEmembership{Senior Member,~IEEE},
        Yong~Niu, Ruisi~He,
        Zhangdui~Zhong,~\IEEEmembership{Senior Member,~IEEE},
        and Zhu~Han,~\IEEEmembership{Fellow,~IEEE}
\thanks{Y. Chen, B. Ai, Y. Niu, R. He and Z. Zhong are with the State Key Laboratory of Rail Traffic Control and Safety, Beijing Engineering Research Center of High-speed Railway Broadband Mobile Communications, and the School of Electronic and Information Engineering, Beijing Jiaotong University, Beijing 100044, China (e-mails:
boai@bjtu.edu.cn, niuy11@163.com; ).}

\thanks{Z. Han is with the University of Houston, Houston, TX 77004 USA (e-mail: zhan2@uh.edu), and also with the Department of Computer Science and Engineering, Kyung Hee University, Seoul, South Korea, 446-701.}

}
\maketitle

\begin{abstract}
Heterogeneous cellular networks (HCNs) with millimeter wave (mm-wave) communications are considered as a promising technology for the fifth generation mobile networks. Mm-wave has the potential to provide multiple gigabit data rate due to the broad spectrum. Unfortunately, additional free space path loss is also caused by the high carrier frequency. On the other hand, mm-wave signals are sensitive to obstacles and more vulnerable to blocking effects. To address this issue, highly directional narrow beams are utilized in mm-wave networks. Additionally, device-to-device (D2D) users make full use of their proximity and share uplink spectrum resources in HCNs to increase the spectrum efficiency and network capacity. Towards the caused complex interferences, the combination of D2D-enabled HCNs with small cells densely deployed and mm-wave communications poses a big challenge to the resource allocation problems. In this paper, we formulate the optimization problem of D2D communication spectrum resource allocation among multiple micro-wave bands and multiple mm-wave bands in HCNs. Then, considering the totally different propagation conditions on the two bands, a heuristic algorithm is proposed to maximize the system transmission rate and approximate the solutions with sufficient accuracies. Compared with other practical schemes, we carry out extensive simulations with different system parameters, and demonstrate the superior performance of the proposed scheme. In addition, the optimality and complexity are simulated to further verify effectiveness and efficiency.
\end{abstract}

\section{Introduction}\label{S1}
With the ever-increasing demands on mobile data streams and the number of connected multi-media devices, some industry and academic experts predict the data rate is expected to increase 1000-fold by year 2020 \cite{Qualcomm}. Many efforts have been made to tackle this issue. Heterogeneous cellular networks (HCNs) with small cells densely deployed underlaying the macrocells have shown great potential to increase frequency reuse and system capacities \cite{niu2017energy}. Besides, due to the scarcity of practical radio frequency resources, many mobile network operators have dedicated to exploit new spectrum bands. Among them, millimeter wave (mm-wave) has emerged as a promising candidate for the fifth-generation (5G) mobile network and attracted tremendous attention for huge bandwidth \cite{Future,Ruisi}. It can achieve throughput in the range of gigabits per second. Moreover, there are already several standards defined for indoor wireless personal area networks (WPANs) or wireless local area networks (WLANs) in the mm-wave band, such as ECMA-387 \cite{ECMA387}, IEEE 802.15.3c \cite{IEEE802153c}, and IEEE 802.11ad. In view of the large bandwidth, it is necessary to divide the mm-wave into multiple bands, as well as for cellular communications. Therefore, in order to meet the fast growth of mobile Internet traffic demands, one promising way is to exploit HCNs in multiple micro-wave bands and multiple mm-wave bands.

To efficiently use the available spectrum resources and maintain a desired quality of service at local users, HCNs with small cells densely deployed have been brought into many studies \cite{spaceair,niu2018,FiWi}. Essentially, implementing HCNs decreases the distance between terminals, which results in lower path losses, reduction in battery consumption, increased energy efficiency and spectrum efficiency. Apart from this, in this paper, the combination of cellular network and mm-wave network makes the advantages of the two networks maximized, and the disadvantages are complementary. It turns out that cellular network has more stable and reliable propagation conditions, and is responsible for network control \cite{mmWavesub6}. However, its transmission rate is limited to fail to meet the continuous growing data traffic. Mm-wave network offers huge bandwidth and provides multiple gigabit data rate, while to some extent, much larger distance associated propagation loss is suffered \cite{ai2017indoor,YongNiu}. For example, the free space path loss at 60 GHz band is 28 dB more than that at 2.4 GHz \cite{singh2011interference}. In addition, a dense deployment of the small cells can increase line-of-sight (LOS) probability and compensate for the blockage of mm-wave networks. Considering the different characteristics of both networks, the benefits of the HCNs with many small cells deployed will be obvious. At last, dividing the cellular and mm-wave communications into multiple bands makes our research more practical.

Device-to-device (D2D) communication has emerged as a promising component to further improve spectral efficiency. In the conventional cellular network, cellular users communicate with each other via the central coordinator, such as base stations (BSs). Different from the infrastructure based cellular network, D2D communications allow two closely located users to communicate directly without involving central controllers \cite{jiajia1,jiajia2,jiajia3}. Due to the proximity of D2D users, the benefits such as reduced end-to-end latency and lower energy consumption are reaped. Then, D2D pairs need to share uplink spectrum resources with cellular users or use the resources in mm-wave bands in multi-cell HCNs. Thus, various interference due to the spectrum sharing needs to be considered, and effective interference management is meaningful.

In this paper, we consider a scenario of multi-cell D2D-enabled heterogeneous cellular network aggregating multiple micro-wave bands and multiple mm-wave bands. The challenges of concerned multi-band resource allocation issues have followed, such as the uniqueness of HCNs, differences in  propagation conditions and power gains of cellular and mm-wave networks, complicated intra- and inter-cell interferences caused by co-band cellular links and D2D links, and algorithm design with priority. Considering these together, the optimization problem of D2D communication spectrum resource allocation among micro-wave bands and mm-wave bands is formulated to maximize the metric of system transmission rate \cite{ma2015interference}, which is defined as the sum rate of all cellular users and D2D pairs in all involved cells, $R=\sum\limits_{c\in C}R_{c}+\sum\limits_{d\in \textbf{D}}R_{d}$ ($C$ and $\textbf{D}$ denote the set of all cellular users and D2D pairs respectively, $R$ denotes the transmission rate). To address this problem, we propose a heuristic algorithm, which takes advantage of the mm-wave communications preferentially after taking into account the characteristics of two networks. As a result, the caused interference is effectively managed and the system performance in terms of the total transmission rate is enhanced. The main contributions of the paper can be summarized as follows.

\begin{itemize}
\item In HCNs with small cells densely deployed underlying the conventional macrocells, we outline the system model consisting of multiple micro-wave bands and mm-wave bands for multiple cellular users and D2D pairs.

\item We propose a heuristic algorithm to make full use of the advantages of cellular network and mm-wave network, while, minimizing interference and maximizing the system transmission rate. Then, we show that the algorithm always yields the near-optimal solution with fairly low computational complexity.

\item Through extensive simulations under various system parameters, we evaluate the system performance of the proposed heuristic algorithm compared with other practical schemes. Besides, the optimality and complexity are also analyzed. On average, the proposed algorithm enhances the system performance in terms of total rate than full mm-wave transmission strategy by about $36\%$.
\end{itemize}

The rest of the paper is organized as follows. In Section~\ref{S2}, we summarize the related work. Section~\ref{S3} outlines the system model and formulates a resource optimization problem. We propose an effective and efficient heuristic algorithm in Section~\ref{S4}. Section~\ref{S5} gives the performance evaluation of the proposed scheme in terms of optimality, complexity and comparison with other three schemes under various system parameters. Finally, we conclude this paper in Section~\ref{S6}.

\section{Related Work}\label{S2}

\begin{figure*}[htbp]
\begin{center}
\includegraphics*[width=1.6\columnwidth,height=4.8in]{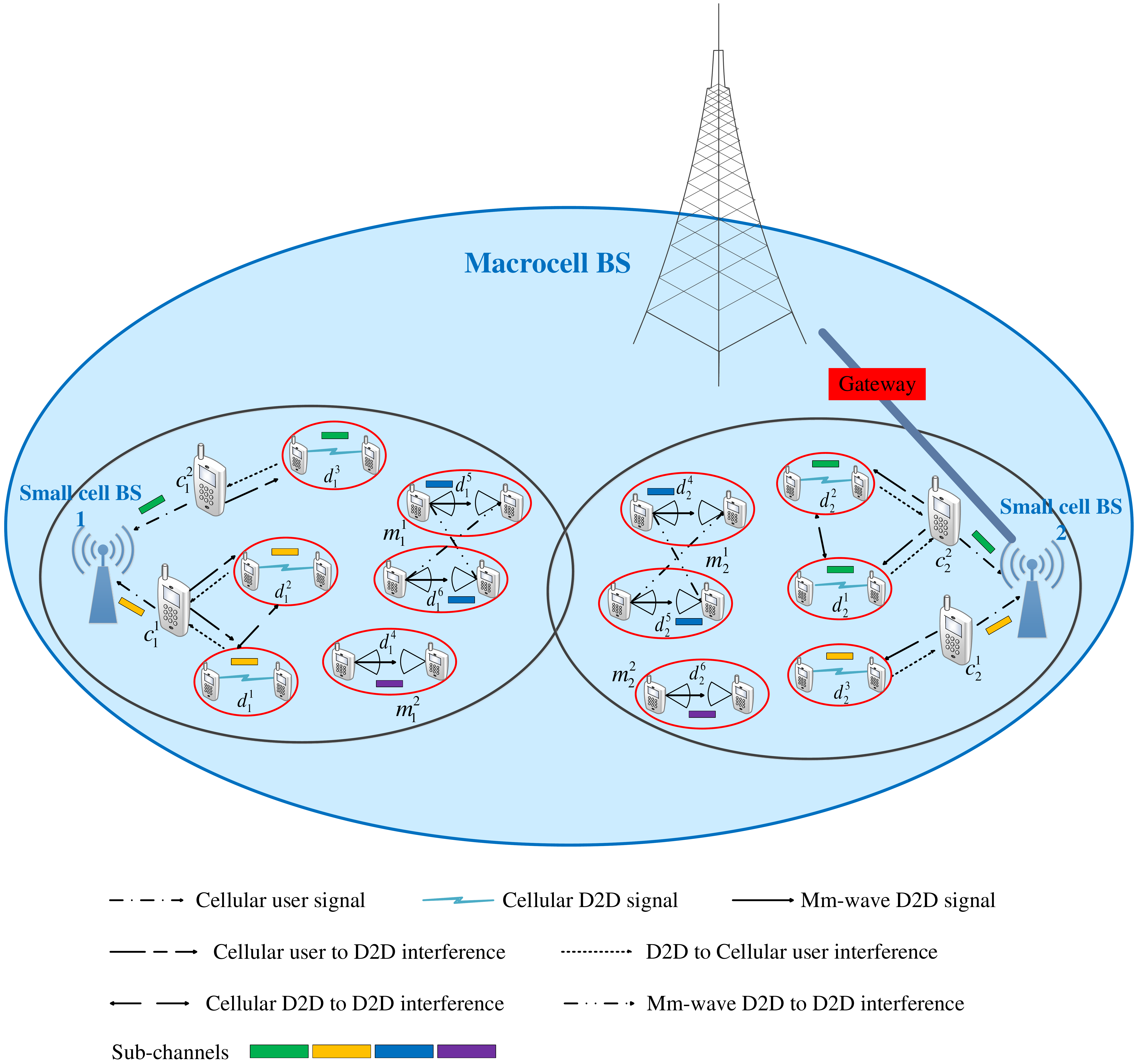}
\end{center}
\caption{Illustration of the resource sharing of D2D communications in a HCN underlaying the macrocell.} \label{fig1}
\end{figure*}
In this section, we partition the related works of resource allocation and interference management into four categories. 1) cellular networks, 2) heterogeneous cellular networks, 3) heterogeneous networks in the mm-wave bands, 4) heterogeneous networks consisting of mm-wave and micro-wave bands.

1) There have been several related works studying power control under a restricted cellular network \cite{ramezani2017joint}, resource allocation for D2D communications under a realistic cellular network \cite{li2014coalitional,xu2013efficiency,Dai}. For example, Ramezani-Kebrya \emph{et al.} \cite{ramezani2017joint} proposed an efficient approximate power control algorithm to maximize the sum rate of a cellular user and a D2D pair with the consideration of the worst-case inter-cell interference (ICI) limit in multiple neighboring cells. To improve spectrum efficiency and enhance system capacity, Li \emph{et al.} \cite{li2014coalitional} proposed a coalition formation game to deal with the interference problem of multiple cellular users and D2D pairs underlaying cellular networks by reasonable resource allocation. Xu \emph{et al.} \cite{xu2013efficiency} proposed an innovative reverse iterative combinatorial auction mechanism to allocate resources to D2D communications underlaying downlink cellular networks. Dai \emph{et al.} \cite{Dai} proposed a framework in which users uploaded data to BSs at most two hops for D2D overlaying multi-channel uplink cellular networks.

2) Different from traditional cellular network, heterogeneous network, which is a multi-cell topology, significantly boosts the overall network capacity. Inspired by it, existing literatures on resource allocation and interference management for heterogeneous cellular networks include \cite{InterferenceCoordination} and \cite{jointaccess}. Wang \emph{et al.} \cite{InterferenceCoordination} proposed a biased cell association scheme with coordinated sub-channel allocation and channel inversion power control for mitigating the co-tier and cross-tier interferences caused by spectrum resource sharing and densification of the small cells in HCNs. Tan \emph{et al.} \cite{jointaccess} proposed a joint access selection and resource allocation scheme to maximize the network capacity in the cache-enabled HCNs with D2D communications.

3) Considering the advantages of mm-wave, such as huge bandwidth, the latest emerging works \cite{zhenxiang,niu2017energy,niu2015,niuyy} open the direction for studying heterogeneous networks in the mm-wave bands. Su \emph{et al.} \cite{zhenxiang} studied the user association and wireless backhaul allocation in a two-tier heterogeneous network operating in the mm-wave band. Niu \emph{et al.} \cite{niu2017energy} developed an energy-efficient mm-wave backhauling scheme to deal with the joint optimization problem of concurrent transmission scheduling and power control of small cells densely deployed in HCNs. Niu \emph{et al.} \cite{niu2015} jointly designed the scheduling problem of radio access and backhaul for small cells in the mm-wave band. Niu \emph{et al.} \cite{niuyy} proposed a coalition formation game based algorithm for optimal sub-channel allocation of access and D2D links in densely deployed multiple mm-wave small cells.

4) Deng \emph{et al.} \cite{mmWavesub6} considered a hierarchical network control framework to address the problems of resource allocation and interference coordination in mm-wave/sub-6 GHz multi-connectivity with relaying scenarios. Chen \emph{et al.} \cite{yali} investigated the D2D communications resource allocation considering single-cell multi-micro-band single-mm-wave-band in HCNs. In this paper, we investigate the scenario of D2D underlaying multi-cell HCNs consisting of multiple micro-wave bands and mm-wave bands, in which resources have to be allocated across frequencies with disparate propagation conditions. The lower micro-wave band is responsible for network control and relatively reliable communications, while mm-wave communications provide high-throughput enhancement. Cellular uplink spectrum resources or mm-wave radio resources are shared by D2D pairs and as a result, intra- and inter-cell interferences are involved and becoming a challenge to the follow-up research. Thus, effective resource allocation and interference handling are the keys to improve the system performance.

Mm-wave bands from 28 GHz to 300 GHz are considered to be a promising candidate for new spectrum in the 5G networks \cite{Zhenyu,Zhenyu2,Guan2017,Zhenyu3,Zhenyu4}. Meanwhile, channel measurements have confirmed some unique characteristics of mm-wave signals, which are different from traditional cellular signals \cite{mmwchannel}. On the one hand, based on large and continuous bandwidth, mm-wave spectrum can offer much higher throughput. On the other hand, mm-wave signal suffers much larger propagation loss due to the high carrier frequency, and it's more vulnerable to blocking and sensitive to obstacles. Consequently, network congestion may happen in mm-wave networks \cite{BlockageRobust}. Another distinguishing characteristic is the directional transmission. The highly directional narrow beams are utilized and the power gains are closely related to the angles of departure/arrival (AoD/AoA). Yu \emph{et al.} \cite{CoverageAnalysis} presented a general framework for the analysis of the coverage probabilities in mm-wave networks, and then conducted a thorough investigation on the impact of directional antenna arrays. Ai \emph{et al.} \cite{ai2017indoor} performed some measurements and simulations on indoor mm-wave massive multiple-input multiple-output (MIMO) channel at 26 GHz. In this paper, integrating mm-wave into HCNs with densely deployed small cells can increase the mm-wave coverage, tackle the problems of path loss and blockage, and further achieve both high capacity and consistent user experience.

\section{System Model and Problem Formulation}\label{S3}

\subsection{System Description}\label{S3-1}
Differing from a simple single-cell scenario, we consider a multi-cell heterogeneous cellular network with multiple potential D2D pairs in this paper. The small cells can be separated or overlapped. In the investigated communication pattern, both the cumulated interference from neighbor cells and the intra-cell interference are considered. Since the HCN is a combination of the cellular network and mm-wave network, we assume two operating modes for each potential D2D pair, the cellular mode and the mm-wave mode. If the D2D pair chooses to operate in the cellular mode, that means it shares the uplink spectrum resource of one cellular user in the same located cell. Otherwise, it uses the radio resource in one of the mm-wave bands. Next, we will discuss system interference in detail.

For the cellular D2D system, the BS is equipped with omnidirectional antennas for cellular communications. Both cellular users and D2D pairs have only one antenna element and its pattern is still omnidirectional, i.e., the gain over all directions is the same. In order to make the complicated interference problem tractable and achieve the maximum spectral efficiency, one cellular user's uplink spectrum resource can be shared simultaneously with multiple D2D pairs. At the same time, the D2D pair is allowed to share no more than one cellular user's spectrum to keep the computational costs low. Besides, the interferences between different cellular sub-channels in the same cell are supposed to be non-existent for the independence of sub-channels. When cellular users of different small cells transmit on the same channel, there exists interference between each other. In general, four kinds of interferences need to be taken into consideration in cellular D2D system, such as cellular link to cellular link interference, cellular link to D2D link interference, D2D link to cellular link interference, and D2D to D2D interference. For the mm-wave D2D system, it is fundamentally different. Highly directional antennas are leveraged for D2D pairs in order to achieve the directional transmission and reception between D2D users in the mm-wave band. Furthermore, significant beam-forming gains are provided. Since there is no forwarding architectures involved, such as BSs, we only need to consider one kind of interference, D2D link to D2D link interference.

In such a system, our goal is to maximize the total transmission rate. To achieve it, a critical challenge should be addressed for efficient D2D communications in HCNs. We need to concentrate on assigning appropriate spectrum resources for all D2D pairs, while satisfying the constraints and mitigating inter- and intra-cell interferences as much as possible. As shown in Fig. \ref{fig1}, we introduce the resource sharing relationship of D2D communications in the HCN underlaying the macrocell, and elaborate on inter- and intra-cell interferences. To make the figure clearer, we simplify the multi-cells into two small cells. The cellular users in each cell are controlled by the corresponding BS, which is connected to the macrocell BS via a direct and high-rate wired connection, called gateway. It should be noted in advance that, 1) the letter $c^{j}_{i}$ represents a cellular user in cell $i$ assigning to sub-channel $j$, 2) the letter $m^{j}_{i}$ represents the mm-wave sub-channel $j$ in cell $i$, 3) the letter $d^{j}_{i}$ represents the D2D $j$ in cell $i$.

For a separate cell, we give detailed intra-cell interference. In the small cell numbered 1, the utilized micro-wave band is divided into two sub-channels. There are two cellular users $c^{1}_{1}$ and $c^{2}_{1}$ occupying them respectively, and 3 D2D pairs with $d^{1}_{1}$ and $d^{2}_{1}$ shared the spectrum resource of $c^{1}_{1}$, $d^{3}_{1}$ shared with $c^{2}_{1}$. Similarly, the available mm-wave band is divided into two sub-channels. $d^{5}_{1}$ and $d^{6}_{1}$ use the mm-wave band in $m^{1}_{1}$, and $d^{4}_{1}$ in $m^{2}_{1}$ band. In addition, a similar network is configured in small cell numbered 2. In each small cell, there are four kinds of interference, 1) cellular user to D2D interference, 2) D2D to cellular user interference, 3) cellular D2D to D2D interference, and 4) mm-wave D2D to D2D interference. Then, the inter-cell interference is also briefly described. There are four sub-channels consisting micro-wave and mm-wave bands in this case, and each sub-channel is represented by a colored rectangle. The links sharing the same sub-channel in the same cell or different cell are marked with the same colored rectangle, that is to say, there must be intra- or inter-cell interferences between any two of them. For example, the received signals at the BS from $c^{1}_{1}$ are interfered by the transmitters of $d^{1}_{1}$, $d^{2}_{1}$, $d^{3}_{2}$ and $c^{1}_{2}$ sharing the same spectrum resource. The signal at the D2D receiver $d^{2}_{2}$ is interfered by the transmitters of $c^{2}_{2}$, $d^{1}_{2}$, $c^{2}_{1}$ and $d^{3}_{1}$. In contrast, in the mm-wave band, the D2D pairs are mutually interfered as they use the same radio resource. The receiver of $d^{4}_{2}$ is interfered by the transmitters of $d^{5}_{1}$, $d^{6}_{1}$ and $d^{5}_{2}$. Regarding to the resource allocation problem to be studied, when all D2D pairs select the cellular mode, since the transmission rate of the cellular link is particularly low, the total system rate will be relatively small. While, when all of them select the mm-wave mode, the interference generated may be serious, and it is not beneficial to the overall performance improvement. Thus, we need to rationally allocate spectrum resources to increase bandwidth utilization and maximize system utility.

\subsection{System Model}\label{S3-2}

\subsubsection{System Model Overview and Assumptions}\label{S3-2-1}

In the system, we assume that there are $n$ small cells, in which multiple cellular users and D2D pairs are uniformly distributed with the BS in the center. The BS associated with cell $i$ is denoted by $b_{i}$. Taking a cell as an example, since the spectrum resources of the HCN are consisted by multiple micro-wave bands and multiple mm-wave bands, for the part of the cellular D2D network, we assume that $N$ cellular bands are defined. Due to the frequency division multiplexing (FDM), we also suppose there are $N$ cellular users occupying different sub-channels without interference between each other. For cell $i$, we denote the set of cellular users as $C_{i}=\{{c}^{1}_{i}, {c}^{2}_{i}, ..., {c}^{N}_{i}\}$. Thus, the set of all cellular users in the system can be expressed as $C=\{C_{1}, C_{2}, C_{3}, ..., C_{n}\}$. In order to be closer to the actual scenario, the number of D2D pairs in each cell is randomly generated. Suppose there are $D_{i}$ D2D pairs, the set of D2D pairs in cell $i$ can be denoted by $\textbf{D}_{i}=\{d^{1}_{i}, d^{2}_{i}, ..., d^{D_{i}}_{i}\}$. Thus, the set of all D2D pairs in the system can be expressed as $\textbf{D}=\{\textbf{D}_{1}, \textbf{D}_{2}, ..., \textbf{D}_{n}\}$. For the part of mm-wave D2D network, we assume that there are $N^{\prime}$ mm-wave bands denoted as $M_{i}=\{m^{1}_{i}, m^{2}_{i}, ..., m^{N^{\prime}}_{i}\}$ in cell $i$. Therefore, the set of mm-wave bands in the system is expressed as $M=\{M_{1}, M_{2}, ... ,M_{n}\}$.

Note that there exists two modes for D2D pair $d_{i}$ to choose. One is to share the uplink resources of cellular users $c_{i}$, and the other is to use mm-wave spectrum $m_{i}$. To better reflect the spectrum resource usage relationship, we define a binary variable $a_{d^{k}_{i}}$ to represent whether the cellular mode or the mm-wave mode is selected by D2D pair $k$ in cell $i$. If the cellular mode is selected, $a_{d^{k}_{i}}=1$. Another case $a_{d^{k}_{i}}=0$ means that the mm-wave spectrum is chosen. For more convenient representation of the relationship of sharing frequency band, we define two other binary variables $x_{c^{j}_{i},c^{j}_{z}}$ and $x_{c^{j}_{i},d^{k}_{i}}$. The former represents the $j$th sub-channel of the cell $i$ is the same as the $j$th of the cell $z$ when it equals to $1$, otherwise, it is not the same. Similarly, the latter represents D2D pair $k$ in cell $i$ shares the spectrum with cellular user $j$ when it equals to $1$. For mm-wave D2D network, we also define two binary variables. One is $x_{m^{y}_{i},m^{y}_{z}}$, representing mm-wave band $y$ of cell $i$ and mm-wave band $y$ of cell $z$ are the same if $x_{m^{y}_{i},m^{y}_{z}}=1$, otherwise, $x_{m^{y}_{i},m^{y}_{z}}=0$. Another is $x_{m^{y}_{i},d^{k}_{i}}$. The variable $x_{m^{y}_{i},d^{k}_{i}}=1$ indicates D2D pair $k$ in cell $i$ occupies the resource of mm-wave band $y$. For all binary variables defined above, we place some reasonable constraints. One is $\sum\limits_{j=1}^{N}x_{c^{j}_{i},d^{k}_{i}}+\sum\limits_{y=1}^{N^{\prime}}x_{m^{y}_{i},d^{k}_{i}}=1, \forall i, k$. In other words, each D2D pair in each cell must share the spectrum of the cellular users or use the radio resource of mm-wave. It cannot transmit directly through the BS. Another one is $\sum\limits_{j=1}^{N}x_{c^{j}_{i},d^{k}_{i}}=a_{d^{k}_{i}}$.

\begin{table}[t]
\begin{center}
\caption{NOTATION SUMMARY}
\begin{tabular}{ll}
\hline
Notation & Description \\
\hline
$n$ & number of small cells  \\
$b_{i}$ & BS of cell $i$ \\
$N$ & number of cellular bands \\
$C_{i}$ & the set of cellular users of cell $i$ \\
$C$ & the set of all cellular users \\
$D_{i}$ & number of D2D pairs of cell $i$ \\
$\textbf{D}_{i}$ & the set of D2D pairs of cell $i$ \\
$D$ & the set of all D2D pairs \\
$N^{\prime}$ & number of mm-wave bands \\
$M_{i}$  & the set of mm-wave bands of cell $i$ \\
$M$ & the set of all mm-wave bands \\
$d^{k}_{i}$ & D2D pair $k$ of cell $i$  \\
$c^{j}_{i}$ & cellular user $j$ of cell $i$  \\
$m^{y}_{i}$ & mm-wave spectrum $y$ of cell $i$  \\
$a_{d^{k}_{i}}$ & mode selection of $d^{k}_{i}$  \\
$x_{c^{j}_{i},c^{j}_{z}}$ & whether $c^{j}_{i}$ and $c^{j}_{z}$ share the same band \\
$x_{c^{j}_{i},d^{k}_{i}}$ & whether $c^{j}_{i}$ and $d^{k}_{i}$ share the same band \\
$x_{m^{y}_{i},m^{y}_{z}}$ & whether $m^{y}_{i}$ and $m^{y}_{z}$ are the same \\
$x_{m^{y}_{i},d^{k}_{i}}$ & whether $d^{k}_{i}$ shares the band of $m^{y}_{i}$  \\
$|h_{0}|^{2}$  & power or second-order statistic of the channel  \\
${\alpha}$   &  path-loss exponent \\
$l_{ij}$  &  distance between $s_{i}$ and $r_{j}$ \\
$P^{c}_{r}(j,i)$  & received power at $r_{i}$ from $s_{j}$ in cellular system \\
$\Gamma^{c}_{i}$  & received SINR at $r_{i}$ from $s_{i}$ in cellular system \\
$G_{t}$  & transmit antenna gain in cellular system \\
$G_{r}$  & receive antenna gain in cellular system \\
$P^{c}_{int,i}$  & interference power received by $r_{i}$ in cellular system \\
$P^{m}_{r}(j,i)$  & received power at $r_{i}$ from $s_{j}$ in mm-wave system \\
$\Gamma^{m}_{i}$  &  received SINR at $r_{i}$ from $s_{i}$ in mm-wave system \\
$G_{t}(i,j)$  & transmit antenna gain of $s_{i}\rightarrow r_{j}$ in mm-wave system \\
$G_{r}(i,j)$  & receive antenna gain of $s_{i}\rightarrow r_{j}$ in mm-wave system \\
$P^{m}_{int,i}$  & interference power received by $r_{i}$ in mm-wave system \\
\hline
\end{tabular}
\label{table1}
\end{center}
\end{table}

\subsubsection{System Channel Model and SINR Representation}\label{S3-2-3}
To maximize the network performance in terms of the system transmission rate, we should pay more emphasis on the key part of signal to interference plus noise ratio (SINR). Considering the cellular D2D network first, we adopt the Rayleigh channel model for small-scale fading due to shadowing and attenuation, while the distance-based path loss is also considered. For communication link $i$, we denote its sender and receiver by $s_{i}$ and $r_{i}$, respectively. The corresponding channel coefficient of the link $i$ is denoted by $h_{ii}$, which can be written as $h_{ii}=|h_{0}|^{2}\cdot{l}^{-\alpha}_{ii}$ under the free space propagation path-loss model, where $|h_{0}|^{2}$ is the power or second-order statistic of the channel, $l_{ii}$ is the distance between sender $s_{i}$ and receiver $r_{i}$, and $\alpha$ is the path-loss exponent \cite{channelmodel,probability}. $h_{ii}$ represents the uplink channel state. On the contrary, $|h_{0}|^{2}$ is considered as a constant within the BS's coverage area, and $h_{0}$ is a complex Gaussian random variable with zero mean and unit variance. Based on the above channel model and path-loss model, for communication link $i$, we derive the expression of the received power at $r_{i}$ from $s_{i}$ as $P^{c}_{r}(i,i)={|h_{0}|}^2\cdot{G_{t}}\cdot{G_{r}}\cdot{l}^{-\alpha}_{ii}\cdot P_{c}$, where $P_{c}$ is the cellular transmission power, $G_{t}$ is the transmit antenna gain and $G_{r}$ is the receive antenna gain. All of them are fixed value for the sake of tractability. The received SINR at $r_{i}$ from $s_{i}$, denoted by $\Gamma^{c}_{i}$, can be expressed as
\begin{equation}
\Gamma^{c}_{i}=\frac{{|h_{0}|}^2G_{t}G_{r}{l}^{-\alpha}_{ii}{P_{c}}}{P^{c}_{int,i}+N_{0c}W_{c}}, \label{eq1}
\end{equation}
where $P^{c}_{int,i}$ is the interference signal power received by user $r_{i}$. The channel is assumed to experience additive white Gaussian noise. $N_{0c}$ is the noise power spectral density of the cellular networks, and $W_{c}$ is the cellular subcarrier bandwidth.

Considering the mm-wave D2D network, the received power at $r_{i}$ from $s_{i}$ can be written as
\begin{equation}
P^{m}_{r}(i,i)=k_{0}G_{t}(i,i)G_{r}(i,i){l}^{-\alpha}_{ii}P_{m}, \label{eq2}
\end{equation}
where $k_{0}$ is a constant coefficient and proportional to ${(\frac{\lambda}{4\pi})}^2$ ($\lambda$ denotes the wavelength) \cite{REX}. We specify the gain which is different from the setting in cellular D2D networks. The antenna gain of $s_{i}$ pointing at direction of $s_{i}\rightarrow r_{i}$ is denoted by $G_{t}(i,i)$. The antenna gain of $r_{i}$ pointing at direction of $s_{i}\rightarrow r_{i}$ is denoted by $G_{r}(i,i)$. Both of them are related with the angles of AoDs/AoAs. $P_{m}$ is the mm-wave transmission power.
There inevitably exists interference between two mutually independent mm-wave communication links $i$ and $j$. Thus, the received interference at $r_{i}$ from $s_{j}$ can be calculated as
\begin{equation}
P^{m}_{r}(j,i)={\rho}k_{0}G_{t}(j,i)G_{r}(j,i){l}^{-\alpha}_{ji}P_{m}, \label{eq3}
\end{equation}
where $\rho$ denotes the multi-user interference (MUI) factor related to the cross correlation of signals from different links.

Combining useful received power, interference, and noise power, we can obtain the received SINR at $r_{i}$ as follows.
\begin{equation}
\Gamma^{m}_{i}=\frac{P^{m}_{r}(i,i)}{{P^{m}_{int,i}}+N_{0m}W_{m}}, \label{eq4}
\end{equation}
where $P^{m}_{int,i}$ is the interference signal power received by user $r_{i}$, the noise onesided power spectral density in the mm-wave D2D network is symbolized by $N_{0m}$, and $W_{m}$ is the bandwidth of mm-wave communication.

\subsubsection{System Transmission Rate}\label{S3-2-3}

In the case of cellular D2D communication, we denote the transmit and receive antenna gain of user equipments and BS as $G_{0}$ and $G_{b}$, respectively. For simplicity, we take them to fixed and reasonable value. Then, we analyze the interference experienced by cellular users and D2D pairs in each cell, and further obtain the uplink transmission rate of each individual contributed to the system. The cumulative interference of the cellular link receiver, which is the BS, partly comes from D2D pairs occupying the same spectrum resource with the cellular user in the same cell, and the remaining comes from cellular users and D2D pairs sharing the same frequency band in other cells. In summary, the interference power at BS $b_{i}$ for cellular user $c^{j}_{i}$ in the small cell $i$ can be expressed as
\begin{equation}
\begin{aligned}
P_{int,c^{j}_{i}}&=\sum\limits_{d^{k}_{i} \in \textbf{D}_{i}}x_{c^{j}_{i},d^{k}_{i}}|h_{0}|^{2}G_{0}G_{b}l^{-\alpha}_{d^{k}_{i},b_{i}}P_{c} \\
&+\sum\limits_{c^{j}_{z}\in C_{z},z \neq i}x_{c^{j}_{i},c^{j}_{z}}(|h_{0}|^{2}G_{0}G_{b}l^{-\alpha}_{c^{j}_{z},b_{i}}P_{c} \\
&+\sum\limits_{d^{k}_{z} \in \textbf{D}_{z}}x_{c^{j}_{z},d^{k}_{z}}|h_{0}|^{2}G_{0}G_{b}l^{-\alpha}_{d^{k}_{z},b_{i}}P_{c}). \label{eq5}
\end{aligned}
\end{equation}
According to the Shannon theory, the achievable transmission rate in bit/s of the cellular user $c^{j}_{i}$, denoted by $R_{c^{j}_{i}}$, can be expressed as
\begin{equation}
R_{c^{j}_{i}}=W_{c}{\log}_2\left(1+\frac{|h_{0}|^{2}G_{0}G_{b}l^{-\alpha}_{c^{j}_{i},b_{i}}P_{c}}{P_{int,c^{j}_{i}}+N_{0c}W_{c}}\right).
\label{eq6}
\end{equation}

Taking D2D pair $k$ in cell $i$ as an example, which is denoted as $d^{k}_{i}$. The received cumulative interference signal of the receiver is from the cellular user $c^{j}_{i}$ and the other D2D pairs sharing the same spectrum resource of $c^{j}_{i}$ in cell $i$, and cellular users and D2D pairs occupying the same sub-channel in other cells. Therefore, the calculation formula of the interference power for the receiver of D2D $d^{k}_{i}$, denoted by $P^{c}_{int,d^{k}_{i}}$, is
\begin{equation}
\begin{aligned}
P^{c}_{int,d^{k}_{i}}&=\sum\limits_{c^{j}_{i} \in C_{i}}x_{c^{j}_{i},d^{k}_{i}}(|h_{0}|^{2}G^{2}_{0}l^{-\alpha}_{c^{j}_{i},d^{k}_{i}}P_{c}\\
&+\sum\limits_{d^{k^{\prime}}_{i} \in \textbf{D}_{i} \backslash \{d^{k}_{i}\}}x_{c^{j}_{i},d^{k^{\prime}}_{i}}|h_{0}|^{2}G^{2}_{0}l^{-\alpha}_{d^{k^{\prime}}_{i},d^{k}_{i}}P_{c})\\
&+\sum\limits_{c^{j}_{i} \in C_{i}}x_{c^{j}_{i},d^{k}_{i}}\sum\limits_{c^{j}_{z} \in C_{z}, z \neq i}x_{c^{j}_{i},c^{j}_{z}}(|h_{0}|^{2}G^{2}_{0}l^{-\alpha}_{c^{j}_{z},d^{k}_{i}}P_{c}\\
&+\sum\limits_{d^{k}_{z} \in \textbf{D}_{z}}x_{c^{j}_{z},d^{k}_{z}}|h_{0}|^{2}G^{2}_{0}l^{-\alpha}_{d^{k}_{z},d^{k}_{i}}P_{c}).
\label{eq7}
\end{aligned}
\end{equation}
With the interference power, we can get the SINR for the receiver of D2D pair $d^{k}_{i}$, denoted by $\Gamma^{c}_{d^{k}_{i}}$, as follows.
\begin{equation}
\Gamma^{c}_{d^{k}_{i}}=\frac{|h_{0}|^{2}G^{2}_{0}l^{-\alpha}_{d^{k}_{i},d^{k}_{i}}P_{c}}{P^{c}_{int,d^{k}_{i}}+N_{0c}W_{c}}.\label{eq8}
\end{equation}
Thus, the transmission rate of the D2D pair $d^{k}_{i}$, denoted by $R^{c}_{d^{k}_{i}}$, is expressed as
\begin{equation}
R^{c}_{d^{k}_{i}}=W_{c}{\log}_2\left(1+\frac{|h_{0}|^{2}G^{2}_{0}l^{-\alpha}_{d^{k}_{i},d^{k}_{i}}P_{c}}{P^{c}_{int,d^{k}_{i}}+N_{0c}W_{c}}\right).\label{eq9}
\end{equation}

In the case of mm-wave D2D communication, interference is more complicated. Different from the single-cell scenario, each D2D pair in the multi-cell scenario suffers interference from all D2D pairs sharing with the same spectrum, not only in the located cell, but also other cells. Thus, we can get the interference power for the receiver of D2D $d^{k}_{i}$, denoted by $P^{m}_{int,d^{k}_{i}}$, as follows.
\begin{equation}
\begin{aligned}
P^{m}_{int,d^{k}_{i}}&=\sum\limits_{m^{y}_{i} \in M_{i}}x_{m^{y}_{i},d^{k}_{i}}\sum\limits_{d^{k^{\prime}}_{i} \in \textbf{D}_{i} \backslash \{d^{k}_{i}\}}x_{m^{y}_{i},d^{k^{\prime}}_{i}}\rho k_{0}\cdot \\
&G_{t}(d^{k^{\prime}}_{i}, d^{k}_{i})G_{r}(d^{k^{\prime}}_{i}, d^{k}_{i})l^{-\alpha}_{d^{k^{\prime}}_{i},d^{k}_{i}}P_{m} \\
&+\sum\limits_{m^{y}_{i} \in M_{i}}x_{m^{y}_{i},d^{k}_{i}}\sum\limits_{m^{y}_{j} \in M_{j}, j \neq i }x_{m^{y}_{i},m^{y}_{j}}\sum\limits_{d^{k}_{j}\in \textbf{D}_{j}}x_{m^{y}_{j},d^{k}_{j}}\rho k_{0}\cdot \\
&G_{t}(d^{k}_{j},d^{k}_{i})G_{r}(d^{k}_{j},d^{k}_{i})l^{-\alpha}_{d^{k}_{j},d^{k}_{i}}P_{m}.
\label{eq10}
\end{aligned}
\end{equation}
Similarly, the expression of the SINR of the D2D receiver $d^{k}_{i}$, denoted by $\Gamma^{m}_{d^{k}_{i}}$, is shown as follows.
\begin{equation}
\Gamma^{m}_{d^{k}_{i}}=\frac{k_{0}G_{t}(d^{k}_{i},d^{k}_{i})G_{r}(d^{k}_{i},d^{k}_{i})l^{-\alpha}_{d^{k}_{i},d^{k}_{i}}P_{m}}{P^{m}_{int,d^{k}_{i}}+
N_{0m}W_{m}}.
\label{eq11}
\end{equation}
Thus, the achievable data rate for the D2D pair $d^{k}_{i}$ in mm-wave band, denoted by $R^{m}_{d^{k}_{i}}$, is given by
\begin{equation}
R^{m}_{d^{k}_{i}}=W_{m}{\log}_2(1+\Gamma^{m}_{d^{k}_{i}}).\label{eq12}
\end{equation}

Combining $R^{c}_{d^{k}_{i}}$ in the cellular D2D network and $R^{m}_{d^{k}_{i}}$ in the mm-wave D2D network, we can obtain the transmission rate of D2D pair $d^{k}_{i}$ in the heterogeneous cellular network system, denoted by $R_{d^{k}_{i}}$, as
\begin{equation}
R_{d^{k}_{i}}=a_{d^{k}_{i}}R^{c}_{d^{k}_{i}}+(1-a_{d^{k}_{i}})(1-P_{out: d^{k}_{i},d^{k}_{i}})R^{m}_{d^{k}_{i}},\label{eq13}
\end{equation}
where $P_{out: d^{k}_{i},d^{k}_{i}}$ denotes the probability of blockage in the line of sight (LOS) path between the sender and the receiver of D2D pair $d^{k}_{i}$ in mm-wave band \cite{Lee2016Connectivity}. The probability of blockage is mainly added to better reflect the characteristics of the mm-wave link, such as mm-wave links are easily blocked by various obstacles. It can be expressed as $P_{out:i,j}=1-e^{-\beta{l}_{ij}}$, where $l_{ij}$ is the distance between user equipments $i$ and $j$, and $\beta$ is the parameter used to reflect the density and size of obstacles, which result in an interruption caused by blockage.

By the rate formula of a single cellular user in (\ref{eq9}) and the rate formula of a single D2D pair in (\ref{eq13}), the total system
rate can be calculated as
\begin{equation}
R=\sum\limits_{i=1}^{n}\sum\limits_{j=1}^{N}R_{c^{j}_{i}}+\sum\limits_{i=1}^{n}\sum\limits_{k=1}^{D_{i}}R_{d^{k}_{i}}.\label{eq14}
\end{equation}

\subsection{Problem Formulation}\label{S3-3}

From the rate formula given in (\ref{eq14}), the system transmission rate is only relevant to the binary variables we defined, such as $x_{c^{j}_{i},c^{j}_{z}}, x_{c^{j}_{i},d^{k}_{i}}, x_{m^{y}_{i},m^{y}_{z}}, x_{m^{y}_{i},d^{k}_{i}}$, and $a_{d^{k}_{i}}$. In view of the relation of equality between $\sum\limits_{j=1}^{N}x_{c^{j}_{i},d^{k}_{i}}$ and $a_{d^{k}_{i}}$, we only take the first four variables into consideration in the analysis of the system transmission rate. For simplicity, we design a matrix ${\mathbf X}$ to represent the spectrum resource sharing relationship. Thus, (\ref{eq14}) can be simplified as a function, denoted by $R({\mathbf X})$. In general, based on the above analysis, the optimization problem of D2D communication resource allocation in multi-cell multi-band heterogeneous cellular networks can be expressed as follows. The goal of the optimization problem is to maximize the system transmission rate and significantly improve system performance.
\begin{align}
&\max\,\,R({\mathbf X})  \notag \\
&s.t.\quad
\begin{cases}
x_{c^{j}_{i},d^{k}_{i}}\in\{0,1\},  \  \forall{c^{j}_{i}}\in{ C_{i}}, {d^{k}_{i}}\in{\textbf{D}_{i}}, 1 \leq i \leq n,\\
 \ \ \ \ \ \ \ \ \ \ \ \ \ \ \ \ \ \ \ \ 1 \leq j \leq N, 1 \leq k \leq D_{i};\\
x_{m^{y}_{i},d^{k}_{i}}\in\{0,1\},  \  \forall{m^{y}_{i}}\in{ M_{i}}, {d^{k}_{i}}\in{\textbf{D}_{i}}, 1 \leq i \leq n,\\
 \ \ \ \ \ \ \ \ \ \ \ \ \ \ \ \ \ \ \ \ 1 \leq y \leq N^{\prime}, 1 \leq k \leq D_{i};\\
x_{c^{j}_{i},c^{j}_{z}}\in\{0,1\},  \  \forall{c^{j}_{i}}\in{ C_{i}}, {c^{j}_{z}}\in{ C_{z}}, 1 \leq i, z \leq n,\\
 \ \ \ \ \ \ \ \ \ \ \ \ \ \ \ \ \ \ \ \ 1 \leq j \leq N;\\
x_{m^{y}_{i},m^{y}_{z}}\in\{0,1\},  \  \forall{m^{y}_{i}}\in{ M_{i}}, {m^{y}_{z}}\in{ M_{z}}, 1 \leq i, z \leq n,\\
 \ \ \ \ \ \ \ \ \ \ \ \ \ \ \ \ \ \ \ \ 1 \leq y \leq N^{\prime};\\
\sum\limits_{j=1}^{N}x_{c^{j}_{i},d^{k}_{i}}+\sum\limits_{y=1}^{N^{\prime}}x_{m^{y}_{i},d^{k}_{i}}=1, \  \forall i,k; \\
\sum\limits_{j=1}^{N}x_{c^{j}_{i},d^{k}_{i}}=a_{d^{k}_{i}}.
\label{eq15}
\end{cases}
\end{align}

Obviously, the formulated optimization problem is considered to be a non-linear integer programming problem. The characteristic of non-linear can be easily seen from the function. In addition, all involved variables are taken values 0 or 1, and both of them are integer. This problem is NP-complete and it is more difficult to solve compared with the 0-1 Knapsack problem \cite{pisinger2005hard}. From the defined function in (\ref{eq15}), we determine the optimization problem and the goal we intend to achieve. Table \ref{table1} summarizes the notations adopted in this section.

\section{Heuristic Algorithm}\label{S4}

\subsection{Motivation and Main Idea} \label{S4-1}

To maximize the overall transmission rate, it is important to come up with an effective and efficient resource allocation scheme so that the spectrum resources are fully utilized. At the same time, the system performance is significantly enhanced.

Without loss of generality, the mm-wave channel transmission rate is about four to five orders of magnitude higher than that of cellular channels. Based on this situation, we first allocate only mm-wave bands to all D2D pairs and leave the cellular sub-channels free. The purpose of this operation is to make the channel with a higher transmission rate can be used preferentially and achieve the optimization goal. Assigning more D2D pairs to share the cellular spectrum resources is of little significance. In order to make the complicated problem tractable, we put all the D2D pairs in the system sharing the same spectrum resource into one set. For example, $N^{\prime}$ carrier frequencies of mm-wave are selected in this paper, and thus all included D2D pairs at the beginning form $N^{\prime}$ sets. The main idea is to perform an appropriate number of switch operations between the $N^{\prime}$ mm-wave sets to preliminarily maximize the system transmission rate. For each set $\Omega$, we define the rate as $R(\Omega)$, which is equivalent to the sum of all D2D rate in the set. It is noted that if the set is sharing the uplink spectrum of cellular users, adding the rate of all cellular users is necessary. After a reasonable allocation of the mm-wave bands, we turn our target to the cellular bands. If the D2D pair in the mm-wave set is exchanged to the cellular set that can make the system rate increased, such switch operation is feasible. By manipulating all the D2D pairs in the mm-wave set, we can move a part of the D2D into the cellular set, eventually making the interferences caused in the mm-wave set and the system transmission rate reach a tradeoff.

\begin{algorithm}[t]
\caption{The Heuristic Algorithm for D2D Pairs Resource Allocation}
\label{alg1}
\begin{algorithmic}[1]
\REQUIRE
Given the number of D2D pairs of cell $i, D_{i}, 1 \leq i \leq n$;\\
Given the initial resource allocation $F^{i}_{ini}=\{c^{i,1}_{ini}, c^{i,2}_{ini}, ..., c^{i,N}_{ini}, m^{i,1}_{ini}, m^{i,2}_{ini}, ..., m^{i,N^{\prime}}_{ini}\}$ of the D2D pairs set $\textbf{D}_{i}$, and $c^{i,1}_{ini}, c^{i,2}_{ini}, ..., c^{i,N}_{ini} = \emptyset$; \\ $c^{j}_{ini}=\{c^{1,j}_{ini}, c^{2,j}_{ini}, ..., c^{n,j}_{ini}\}$, for cellular band $1 \leq j \leq N$; \\
$m^{j}_{ini}=\{m^{1,j}_{ini}, m^{2,j}_{ini}, ..., m^{n,j}_{ini}\}$, for mm-wave band $1 \leq j \leq N^{\prime}$; \\
$F_{ini}=\{c^{1}_{ini}, c^{2}_{ini}, ..., c^{N}_{ini}, m^{1}_{ini}, m^{2}_{ini}, ..., m^{N^{\prime}}_{ini}\}$; \\
$F_{ini} \rightarrow F_{cur}, num1=0, num2=0$.

\REPEAT
\STATE Choose one D2D Pair $k$, and denote its mm-wave set as $m^{j}_{cur}\subset F_{cur}$;
\STATE Uniformly randomly search for another possible mm-wave set $m^{j^{\prime}}_{cur}\subset F_{cur}, m^{j^{\prime}}_{cur}  \neq  m^{j}_{cur}$;
\IF{$R(m^{j}_{cur}\backslash k)+R(m^{j^{\prime}}_{cur} \bigcup k) > R(m^{j}_{cur})+R(m^{j^{\prime}}_{cur})$}
\STATE D2D pair $k$ leaves its set $m^{j}_{cur}$, and joins the new set $m^{j^{\prime}}_{cur}$;\ $num1=0$;
\STATE Update the $F_{cur}$ as $(F_{cur} \backslash \{m^{j}_{cur},m^{j^{\prime}}_{cur}\})\bigcup \{m^{j}_{cur} \backslash \{k\},m^{j^{\prime}}_{cur}\bigcup \{k\}\}\longrightarrow F_{cur}$;
\ELSE
\STATE $num1=num1+1$;
\ENDIF
\UNTIL the procedure converges.
\REPEAT
\STATE Choose one D2D Pair $k$, and denote its mm-wave set as $m^{j}_{cur}\subset F_{cur}$;
\STATE Uniformly randomly search for another possible cellular set $c^{j^{\prime}}_{cur}\subset F_{cur}$;
\IF{$R(m^{j}_{cur}\backslash k)+R(c^{j^{\prime}}_{cur} \bigcup k) > R(m^{j}_{cur})+R(c^{j^{\prime}}_{cur})$}
\STATE D2D pair $k$ leaves its set $m^{j}_{cur}$, and joins the new set $c^{j^{\prime}}_{cur}$;\ $num2=0$;
\STATE Update the $F_{cur}$ as $(F_{cur} \backslash \{m^{j}_{cur},c^{j^{\prime}}_{cur}\})\bigcup \{m^{j}_{cur} \backslash \{k\},c^{j^{\prime}}_{cur}\bigcup \{k\}\}\longrightarrow F_{cur}$;
\ELSE
\STATE $num2=num2+1$;
\ENDIF
\UNTIL the procedure converges.
\ENSURE $F_{cur}\rightarrow F_{fin}$.
\end{algorithmic}
\end{algorithm}

\subsection{Heuristic Algorithm}\label{S4-2}
In this subsection, we describe the proposed algorithm in details. Its pseudo code is shown in Algorithm \ref{alg1}.
Initially, we randomly generate a certain number of D2D pairs for each cell. To make it reasonable, we set an upper bound for the number of D2D in each cell. Next, we give the initial resource allocation $F^{i}_{ini}=\{c^{i,1}_{ini}, c^{i,2}_{ini}, ..., c^{i,N}_{ini}, m^{i,1}_{ini}, m^{i,2}_{ini}, ..., m^{i,N^{\prime}}_{ini}\}$ for cell $i$, where $c^{i,1}_{ini}, c^{i,2}_{ini}, ..., c^{i,N}_{ini} = \emptyset$. Then, we integrate all D2D pairs and classify them sharing the same frequency band into a set. From step 1 to step 10, we complete the optimization between $N^{\prime}$ mm-wave bands. Indeed, it is feasible to select users randomly in step 2. However, in order to allow each D2D pair to participate in this iterative process, we label them and iterate in the order of labels. Knowing the D2D's current located set $m^{j}_{cur}$, we uniformly randomly search for another one $m^{j^{\prime}}_{cur}$. From steps 4 to 9, if we make the system transmission rate increased by exchanging the D2D from $m^{j}_{cur}$ to $m^{j^{\prime}}_{cur}$, the switch operation is performed. What needs to be emphasized is that our decision condition in step 4 only involves two sets, because the rest of the sets have not changed. Thus, the increase in the rate of these two sets means that the system rate increases. If the relation is satisfied, D2D pair $k$ leaves its set $m^{j}_{cur}$, and joins the new set $m^{j^{\prime}}_{cur}$. At the same time, we update the set $F_{cur}$. In addition, we set the corresponding iteration termination condition by defining the parameter $num1$, which is the number of consecutive recordings without switch operations. Thus, when we perform the switch operation, we zero the parameter; when not executed, $num1=num1+1$. From steps 11 to 20, considering the cellular spectrum resources, we make the second improvement in system performance. Similarly, in step 13, we randomly search for another possible cellular set $c^{j^{\prime}}_{cur}$. If switching users to the cellular band is beneficial to system performance, we also perform this operation. Steps 11 to 20 are similar to steps 1 to 10. Parameter $num2$ is another record parameter established to distinguish two iterations. Finally, we return a union of each set of frequency bands.

In Algorithm \ref{alg1}, we set the number of iterations for the two loops from steps 1 to 10 and from steps 11 to 20 to be $N_{1}$ and $N_{2}$, respectively. In each iteration, based on the judgment criterion that whether the system utility will increase, the D2D pair decides to perform a switch operation or not. Thus, there will be at most 1 switch operation in each iteration, and then, the computational complexity lies in the number of total iterations and can be expressed as $O (N_{1}+N_{2})$.

\section{Performance Evaluation}\label{S5}

In this section, we evaluate the performance of the proposed heuristic algorithm under various system parameters. At the same time, the optimality and complexity of the algorithm is also simulated. In addition, we compare our scheme with three other schemes in terms of the system transmission rate. Finally, a detailed analysis of the simulation results is presented.

\begin{table}[t]
\begin{center}
\caption{SIMULATION PARAMETERS}
\begin{tabular}{ccc}
\hline
Parameter & Symbol  & Value \\
\hline
mm-wave bandwidth & $W_{m}$ & 1080 MHz  \\
Cellular carrier bandwidth & $W_{c}$ & 15 KHz \\
mm-wave noise spectral density &  $N_{0m}$  & -134 dBm/MHz \\
Cellular noise spectral density & $N_{0c}$ & -174 dBm/Hz \\
mm-wave transmission power & $P_{m}$ & 20 dBm \\
Cellular transmission power&  $P_{c}$ & 23 dBm \\
Path loss exponent  &  $\alpha$   &  2 \\
MUI factor &  $\rho$  & 1 \\
Half-power beamwidth  &   $\theta_{-3dB}$ &    $30^{\circ}$ \\
Blockage parameter  &   $\beta$   & 0.01 \\
Antenna gains of device &   $G_{0}$  &    0.5 dBi \\
Antenna gains of BS &  $G_{b}$    &  14 dBi \\
\hline
\end{tabular}
\label{table2}
\end{center}
\end{table}

\subsection{Simulation Setup}\label{S5-1}

\begin{figure*}[t]
\begin{minipage}[t]{0.5\linewidth}
\centering
\includegraphics[width=0.8\columnwidth,height=2.3in]{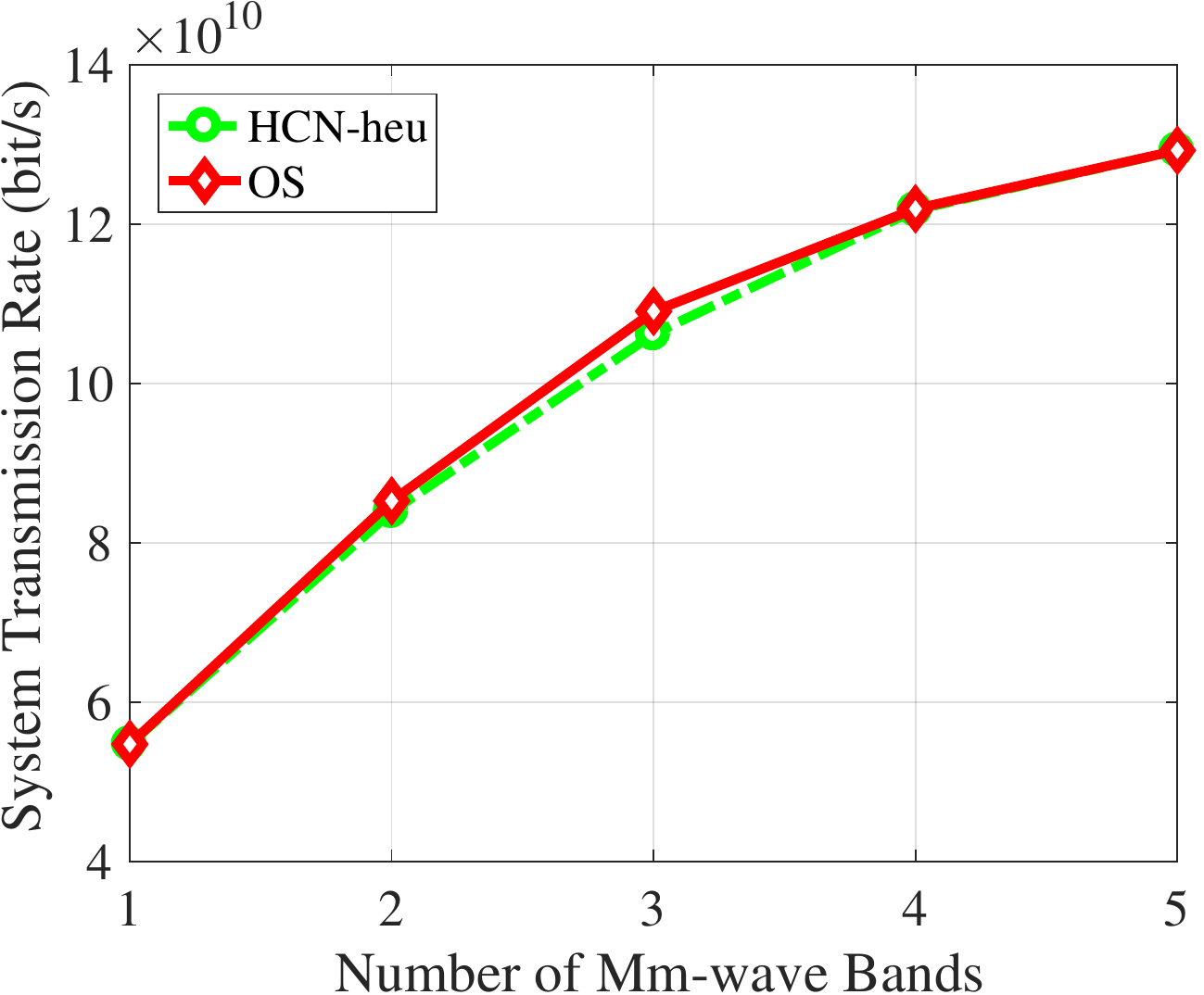}
\centerline{\small (a)}
\end{minipage}%
\begin{minipage}[t]{0.5\linewidth}
\centering
\includegraphics[width=0.8\columnwidth,height=2.3in]{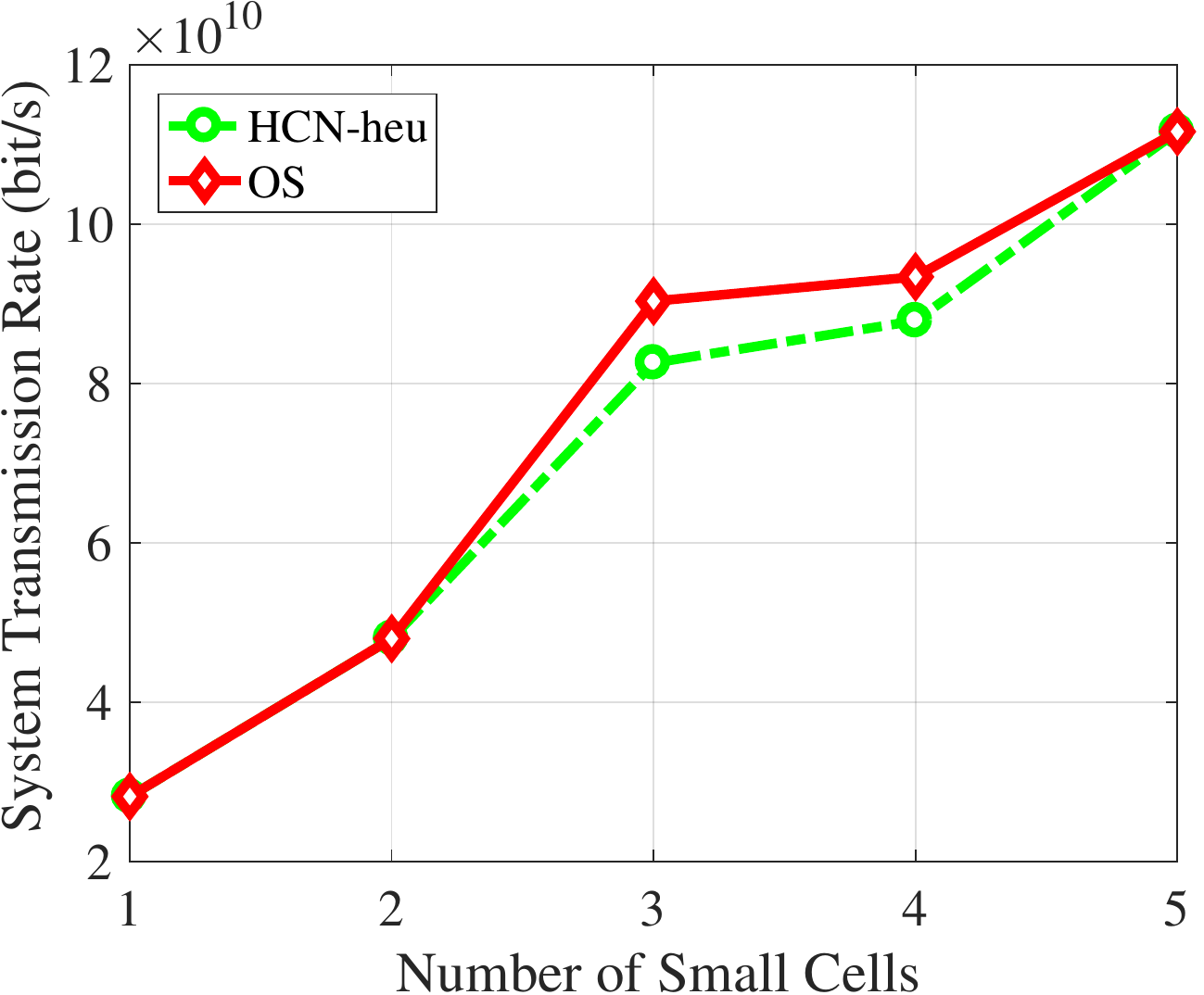}
\centerline{\small (b)}
\end{minipage}
\caption{System transmission rate comparison of HCN-heu and OS with (a) different numbers of mm-wave bands and (b) different numbers of small cells.}
\vspace*{-3mm}
\label{fig2}
\end{figure*}

In the simulation, we consider a heterogeneous cellular network system consisting of multiple small cells. The system is designed as a square area of $100m\times 100m$ and each cell is a circular area with radius $R=20m$. In each cell, both cellular users and D2D pairs are uniformly deployed in the inner circle area, and the BS is located in the center. Obviously, it is likely that there is overlap between different small cells. Without loss of generality, we set the path-loss exponent in free space propagation model considered in this paper as 2. On the one hand, D2D communication is a linear connection channel formed when the physical distance between two users is relatively short. It is more reasonable to set an upper bound on D2D distance. However, since the radius of the small cell is small, there is no additional upper bound setting in this paper. On the other hand, the widely used realistic directional antenna model is adopted in the mm-wave D2D network, which is a main lobe of Gaussian form in linear scale and constant level of side lobes \cite{antennamodel}. It is the reference antenna model with side-lobe for IEEE 802.15.3c. Based on this model, the gain of a directional antenna in units of decibel (dB), denoted by $G(\theta)$, can be expressed as

\begin{equation}
G({\theta})=\left\{
\begin{array}{rcl}
G_{m}-3.01\cdot\left({\frac{2\theta}{\theta_{-3dB}}}\right)^2,&& {0^{\circ}\leq {\theta} \leq {{\theta}_{ml}}/2},\\
G_{sl},&& {{{\theta}_{ml}}/2 \leq {\theta} \leq 180^{\circ}},\\
\end{array} \right. \label{eq16}
\end{equation}
where $\theta$ denotes an arbitrary angle within the range $[0^{\circ},180^{\circ}]$, ${\theta}_{-3dB}$ denotes the angle of the half-power beamwidth, and ${\theta}_{ml}$ denotes the main lobe width in units of degrees. The relationship between ${\theta}_{ml}$ and ${\theta}_{-3dB}$ is ${\theta}_{ml}=2.6\cdot {\theta}_{-3dB}$. $G_{m}$ is the maximum antenna gain, and can be expressed as
\begin{equation}
G_{m}=10\log\left(\frac{1.6162}{\sin(\theta_{-3dB}/2)}\right)^2.
\label{eq17}
\end{equation}
$G_{sl}$ denotes the side lobe gain, which can be obtained by
\begin{equation}
G_{sl}=-0.4111\cdot \ln(\theta_{-3dB})-10.579.
\label{eq18}
\end{equation}
The other simulation parameters are shown in Table \ref{table2}.

In this paper, we focus on the system performance in terms of system transmission rate. To show the superior performance of the proposed heuristic algorithm, we compare it, labeled as \textbf{HCN-heuristic} (HCN-heu) with the following three algorithms:

$a)$ \textbf{Mm-wave Communication} (MMW), where each D2D pair in the system can only choose the spectrum resources from one of the mm-wave bands to share. In addition, all cellular users occupy the sub-channels individually, with no interference between each other, and no interference from D2D pairs.

$b)$ \textbf{HCN} (HCN), where the system spectrum resources include multiple cellular carrier frequencies and multiple mm-wave carrier frequencies. In other words, the system can randomly assign D2D pairs to share one cellular user's uplink resources or occupy one of the mm-wave sub-channels. The difference between this algorithm and the proposed HCN-heu algorithm is that there is no effective heuristic algorithm for resource allocation to further improve the system performance, but simply allocates randomly selected resources to the D2D pairs.

$c)$ \textbf{Mm-wave One Band} (MMW-1), which allocates the unique mm-wave sub-channel to all D2D pairs. Because of the severe spectrum interference, the advantage of mm-wave communication is greatly weakened. Thus, the link's utility contributed to the system is small. The difference between this algorithm and the MMW is just the number of mm-wave bands.

\subsection{Comparison With Optimal Solution}\label{S5-2}

In this subsection, we perform some simulations and further give insights into the gap between the proposed algorithm and the optimal solution (OS). Since the optimal solution is obtained by the exhaustive search method, the complexity is extremely high. Thus, we set both the number of small cells and the number of cellular bands to be 2, the number of D2D pairs in each cell to be 4, and vary the number of mm-wave bands from 1 to 5 to obtain the simulation results shown in Fig. \ref{fig2}(a). Besides, both the number of cellular bands and mm-wave bands are set to be 1, while varying the number of cells from 1 to 5 to obtain the simulation results shown in Fig. \ref{fig2}(b). From these two figures, we can see the system transmission rate achieved by HCN-heu, shown by the dot and dash curve, has an excellent approximation to that achieved by OS, shown by the solid line curve. In order to quantitatively analyze the approximation of the two curves, we select the average deviation between the results obtained by HCN-heu and OS as an indicator, which is expressed as follows.

\begin{equation}
Average \ Deviation = \frac{1}{5}\sum\limits_{n=1}^{5}{\frac{R_{OS}(n)-R_{HCN-heu}(n)}{R_{OS}(n)}}, \label{eq24}
\end{equation}
where $R_{OS}(n)$ and $R_{HCN-heu}(n)$ denote the system transmission rate obtained by OS and HCN-heu, respectively, with the number of mm-wave bands or small cells $n$. As a result, the average deviation between the HCN-heu and OS is about $1.4\%$ in Fig. \ref{fig2}(a), while the average deviation is about $7.3\%$ in Fig. \ref{fig2}(b). Thus, we complete the demonstration that the proposed heuristic algorithm can obtain a sufficiently accurate solution, which is close to the optimal solution of the optimization problem.

\subsection{Comparison With Other Schemes}\label{S5-3}

\begin{figure}[t]
\begin{center}
\includegraphics*[width=0.8\columnwidth,height=2.3in]{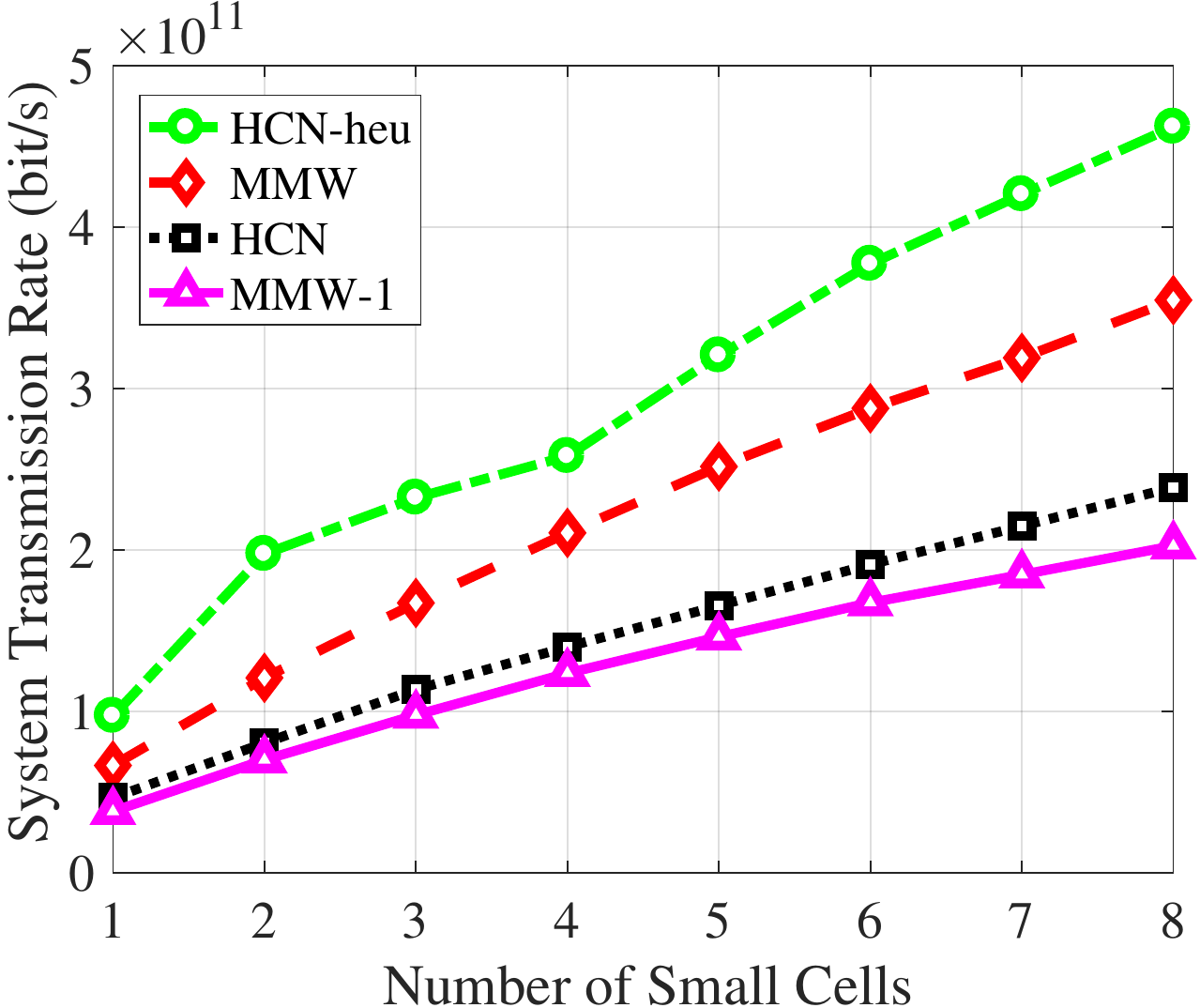}
\end{center}
\caption{System transmission rate comparison of four resource allocation algorithms with different numbers of small cells.}
\label{fig3}
\end{figure}

$1) \ Number \ of \ Small \ Cells:$ In Fig. \ref{fig3}, we set the number of mm-wave bands and the number of cellular bands as 3. Then, we plot the system transmission rate comparison of the four schemes varying the number of small cells from 1 to 8. From the figure, we can see that for all algorithms, the total transmission rate increases with the increase in the number of small cells. This is because the number of cellular users and D2D pairs has increased, and their contributions to the system can completely offset the complex interferences caused to the system. In addition, the proposed algorithm is superior to other practical schemes. At the number of small cells of 8, the system rate achieved by our scheme is higher than MMW and HCN about $30\%$ and $94\%$, respectively. Averagely, the HCN-heu outperforms MMW and HCN about $36\%$ and $103\%$, respectively.

\begin{figure}[t]
\begin{center}
\includegraphics*[width=0.8\columnwidth,height=2.3in]{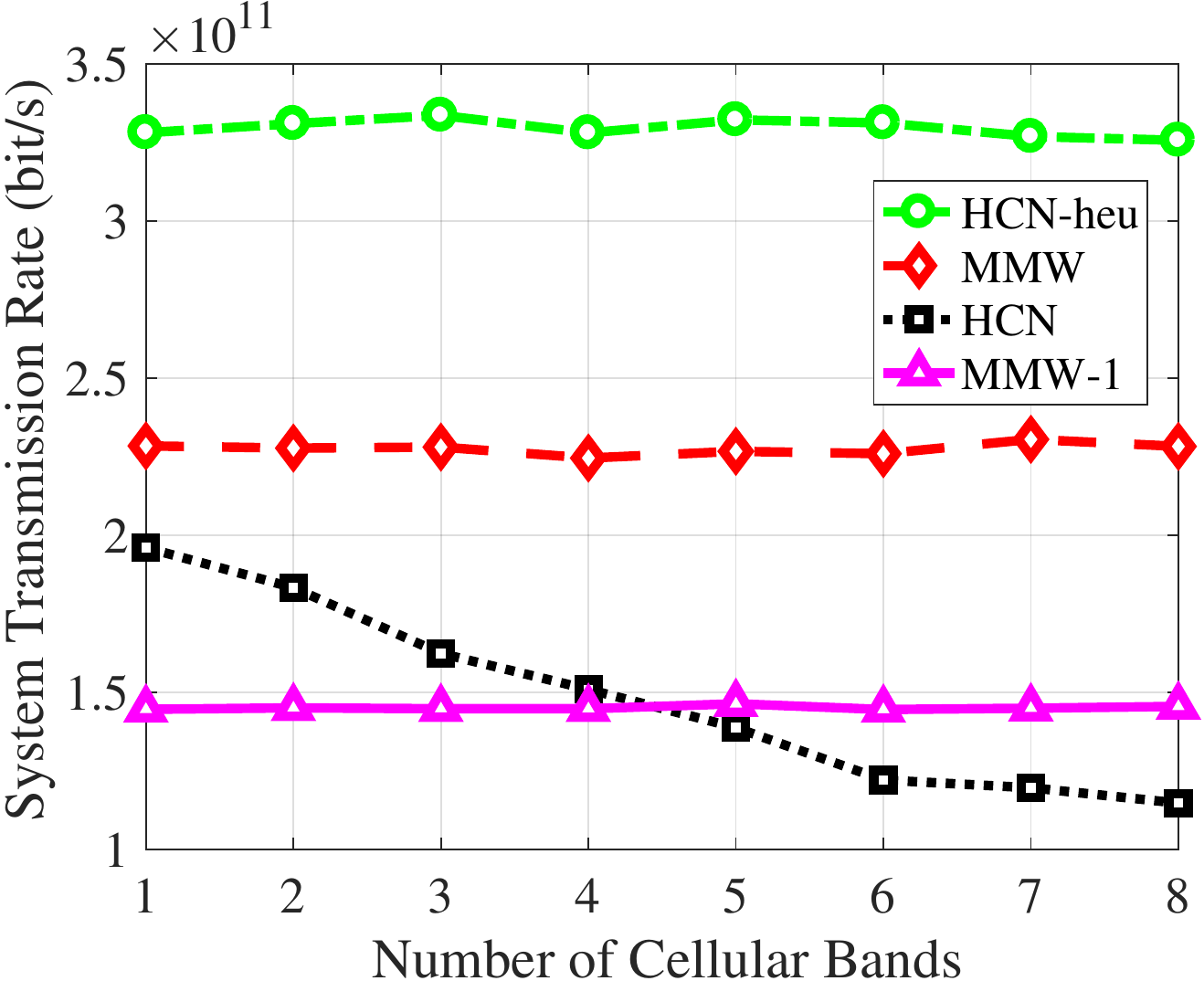}
\end{center} \caption{System transmission rate comparison of four resource allocation algorithms with different numbers of cellular bands.}
\label{fig4}
\end{figure}

$2) \ Number \ of \ Cellular \ Bands:$ In Fig. \ref{fig4}, we set the number of small cells and mm-wave bands to be 5 and 3, respectively. Then, we plot the system transmission rate comparison of the four schemes, varying the number of cellular bands from 1 to 8, where the system rate is in the unit of bit per second. From the simulation results, we can observe that the proposed algorithm achieves the highest system transmission rate and significantly exceeds other ones. As the number of cellular bands changes, the curves of HCN-heu, MMW and MMW-1 do not change substantially, which is because the transmission rate of mm-wave is significantly better than the cellular transmission rate. Thus, most D2D pairs in HCN-heu and all D2D pairs in MMW and MMW-1 choose to use the spectrum resource of mm-wave, and there is no point in increasing the number of cellular bands. The scheme HCN declines due to the fact that some D2D pairs are randomly assigned to the cellular sub-channels with the number of cellular bands increased. At the number of cellular bands of 8, the system rate achieved by the proposed scheme is higher than MMW and HCN about $43\%$ and $92\%$, respectively.

\begin{figure}[t]
\begin{center}
\includegraphics*[width=0.8\columnwidth,height=2.3in]{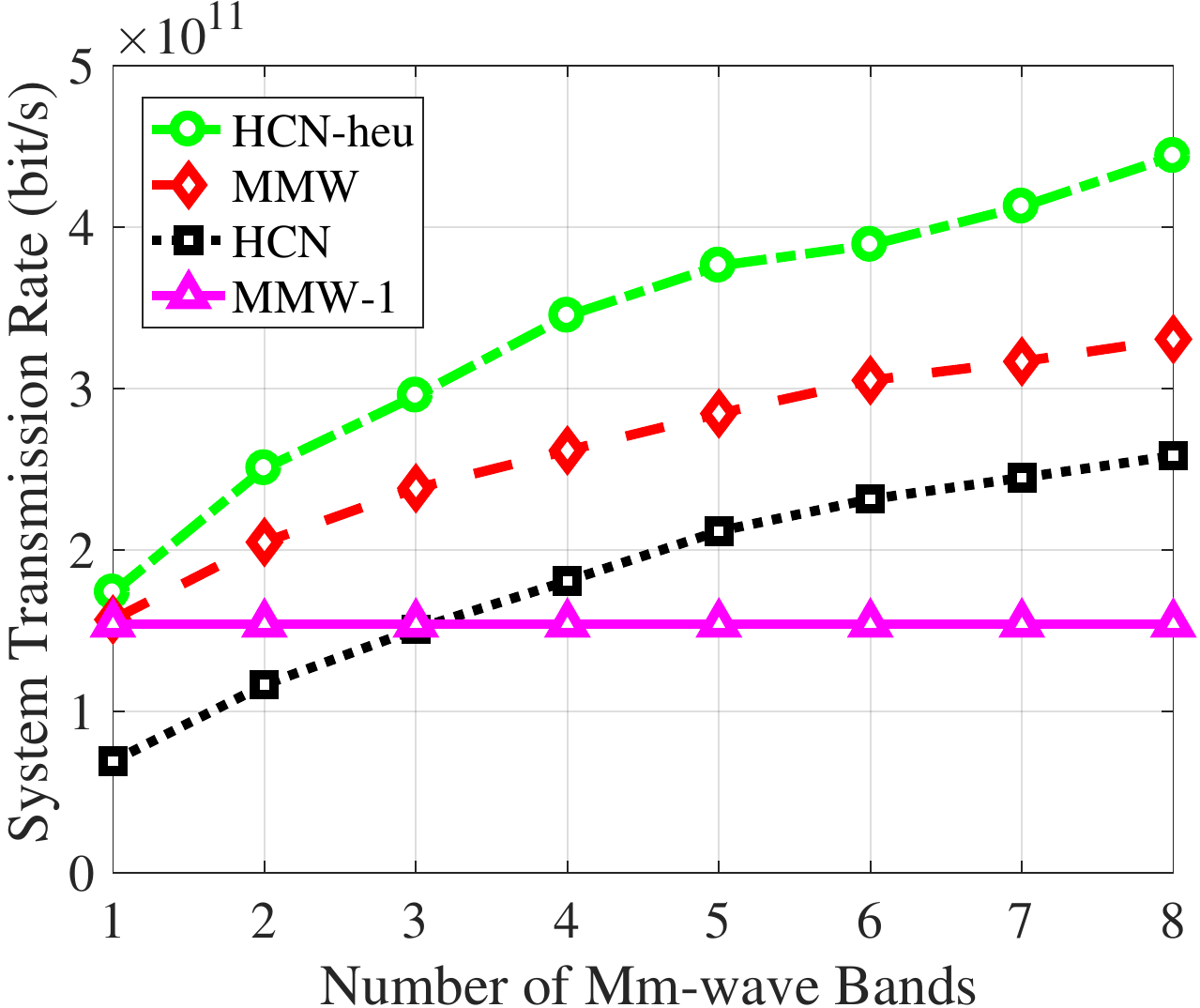}
\end{center} \caption{System transmission rate comparison of four resource allocation algorithms with different numbers of mm-wave bands.}
\label{fig5}
\end{figure}

\begin{figure}[t]
\begin{center}
\includegraphics*[width=0.8\columnwidth,height=2.3in]{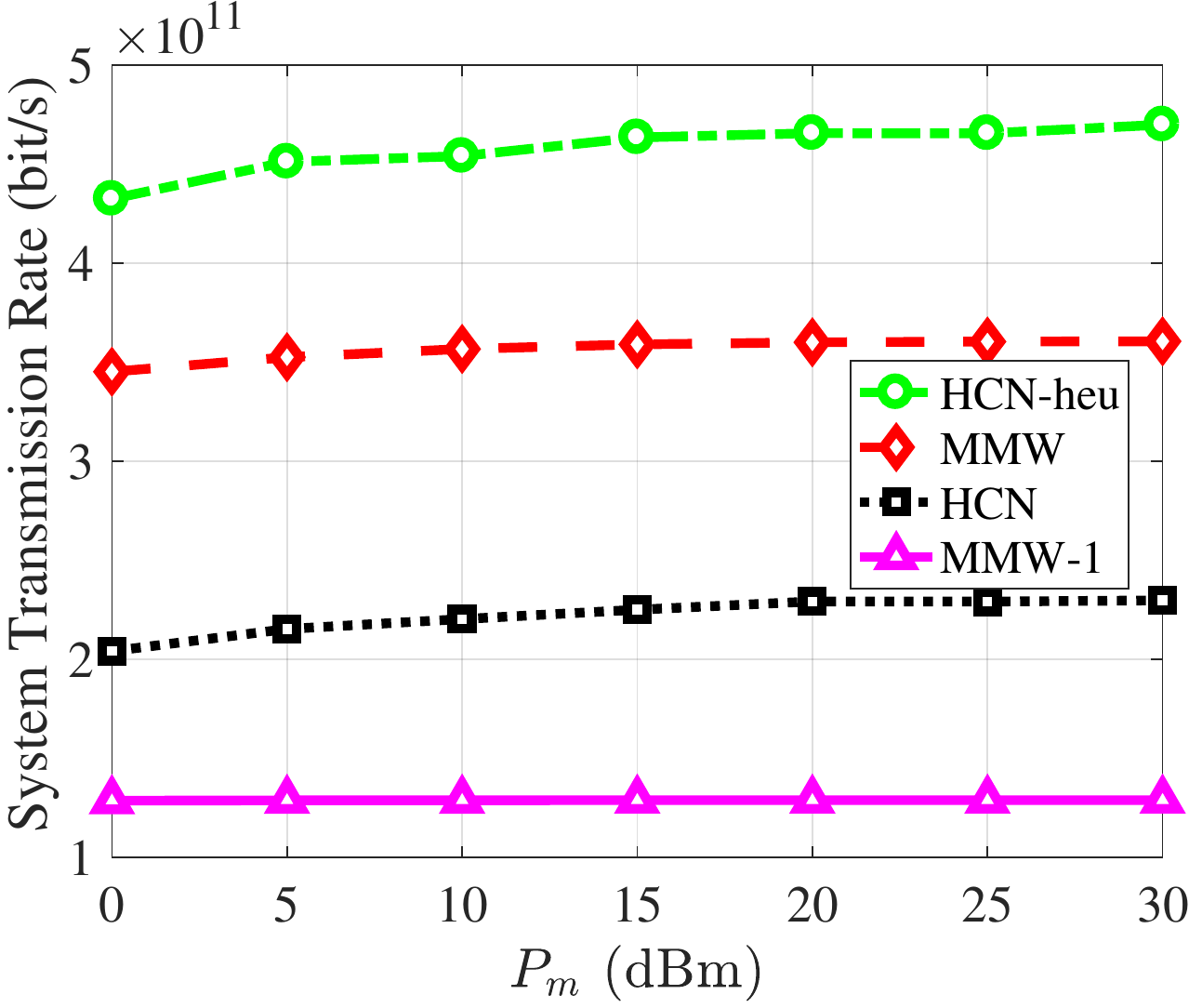}
\end{center} \caption{System transmission rate comparison of four resource allocation algorithms with different $P_{m}$.}
\label{fig6}
\end{figure}

$3) \ Number \ of \ Mm-wave \ Bands:$ In Fig. \ref{fig5}, we set the number of small cells and cellular bands to be 5 and 3, respectively. Then, we plot the achieved system rate of four schemes under different numbers of mm-wave bands. Comparing with other schemes, the proposed heuristic algorithm once again shows a good advantage. Indeed, the mm-wave band has a huge contribution to the entire system. At the same time, with the number of mm-wave bands increasing, D2D pairs have more choices to access one of the mm-wave frequencies in algorithms HCN-heu, MMW and HCN involving mm-wave. Consequently, both the interference power in the cellular frequency set and the mm-wave frequency set are reduced. In summary, these three curves show an upward trend. When the number of mm-wave bands is equal to 8, the system transmission rate of HCN-heu is larger than that of MMW and HCN about $34\%$ and $72\%$, respectively. Besides, the scheme MMW-1 is unchanged because there is only one mm-wave sub-channel.

\begin{figure}[t]
\begin{center}
\includegraphics*[width=0.8\columnwidth,height=2.3in]{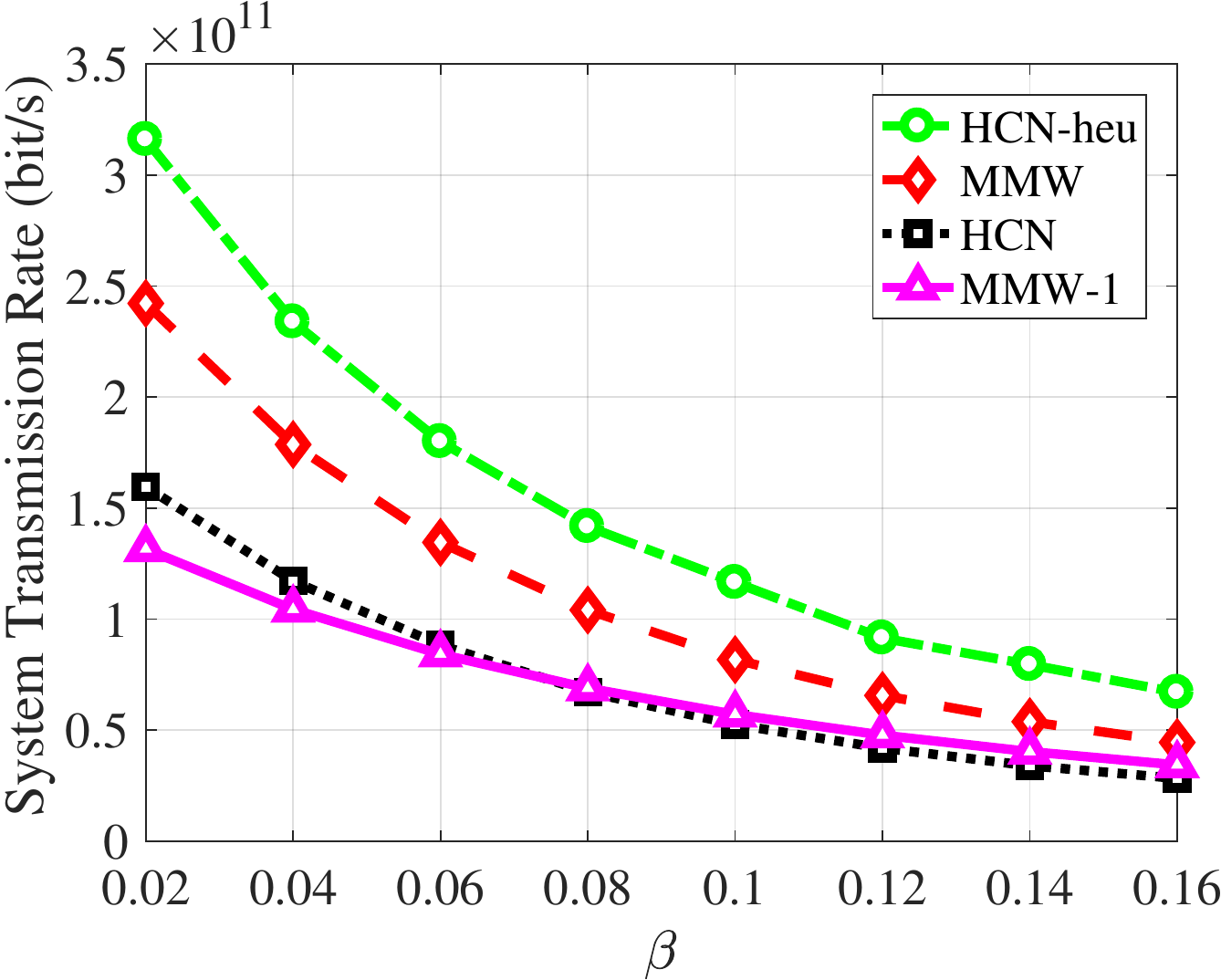}
\end{center} \caption{System transmission rate comparison of four resource allocation algorithms with different $\beta$.}
\label{fig7}
\end{figure}

\begin{figure}[t]
\begin{center}
\includegraphics*[width=0.8\columnwidth,height=2.3in]{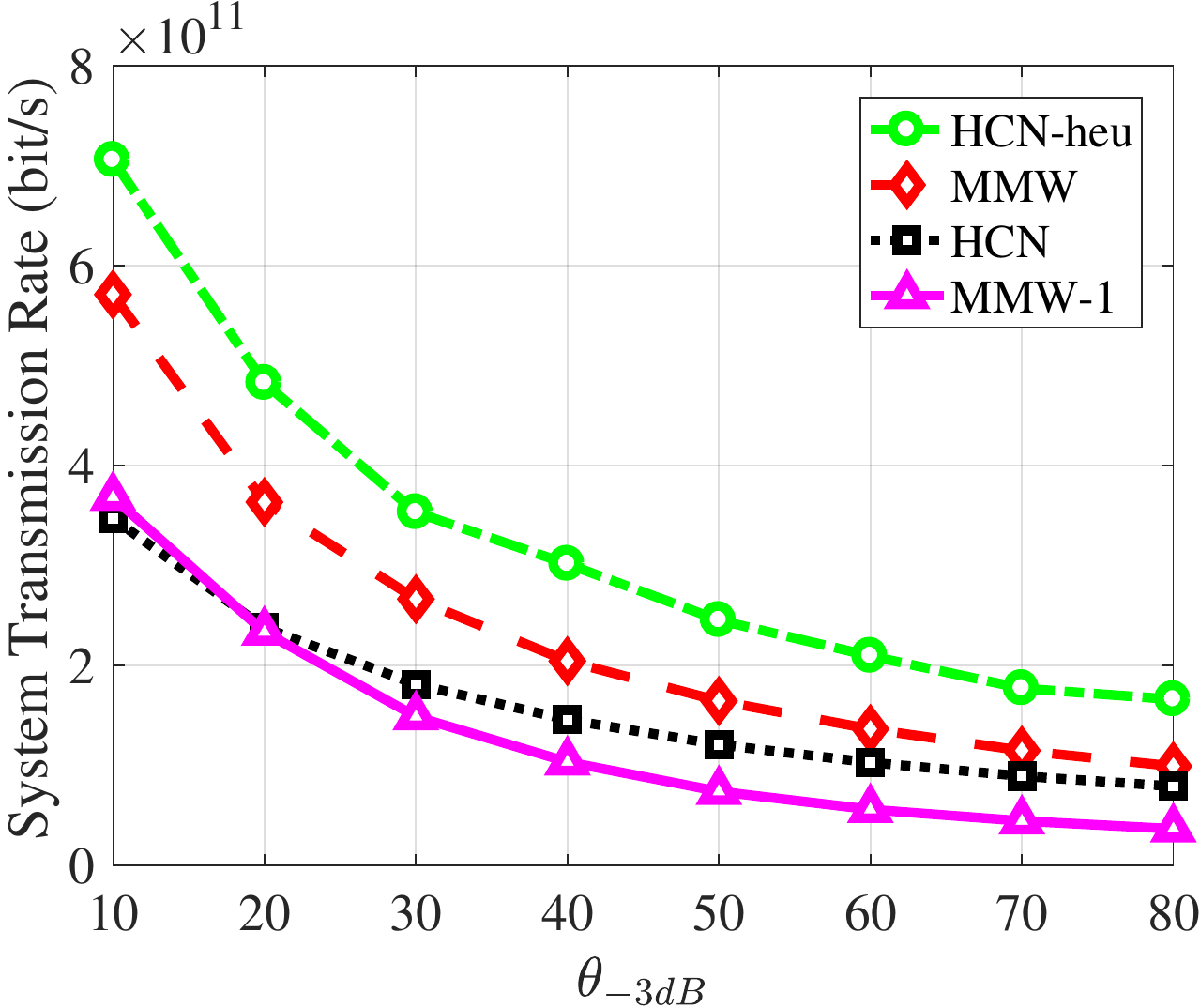}
\end{center} \caption{System transmission rate comparison of four resource allocation algorithms with different $\theta_{-3dB}$.}
\label{fig8}
\end{figure}

\begin{figure}[htbp]
\begin{center}
\includegraphics*[width=0.8\columnwidth,height=2.3in]{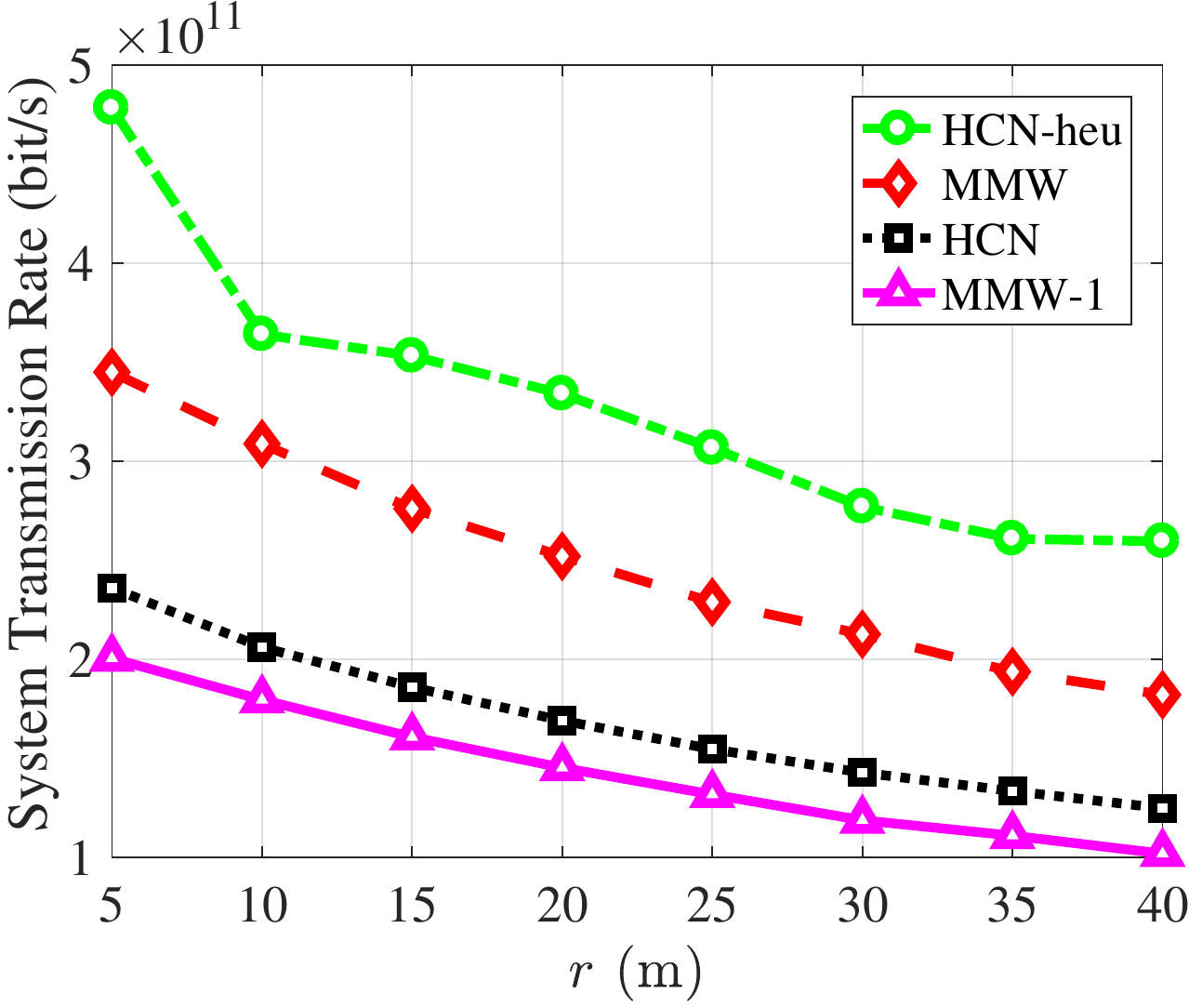}
\end{center} \caption{System transmission rate comparison of four resource allocation algorithms with different $r$.}
\label{fig9}
\end{figure}

\begin{figure}[htbp]
\begin{center}
\includegraphics*[width=0.8\columnwidth,height=2.3in]{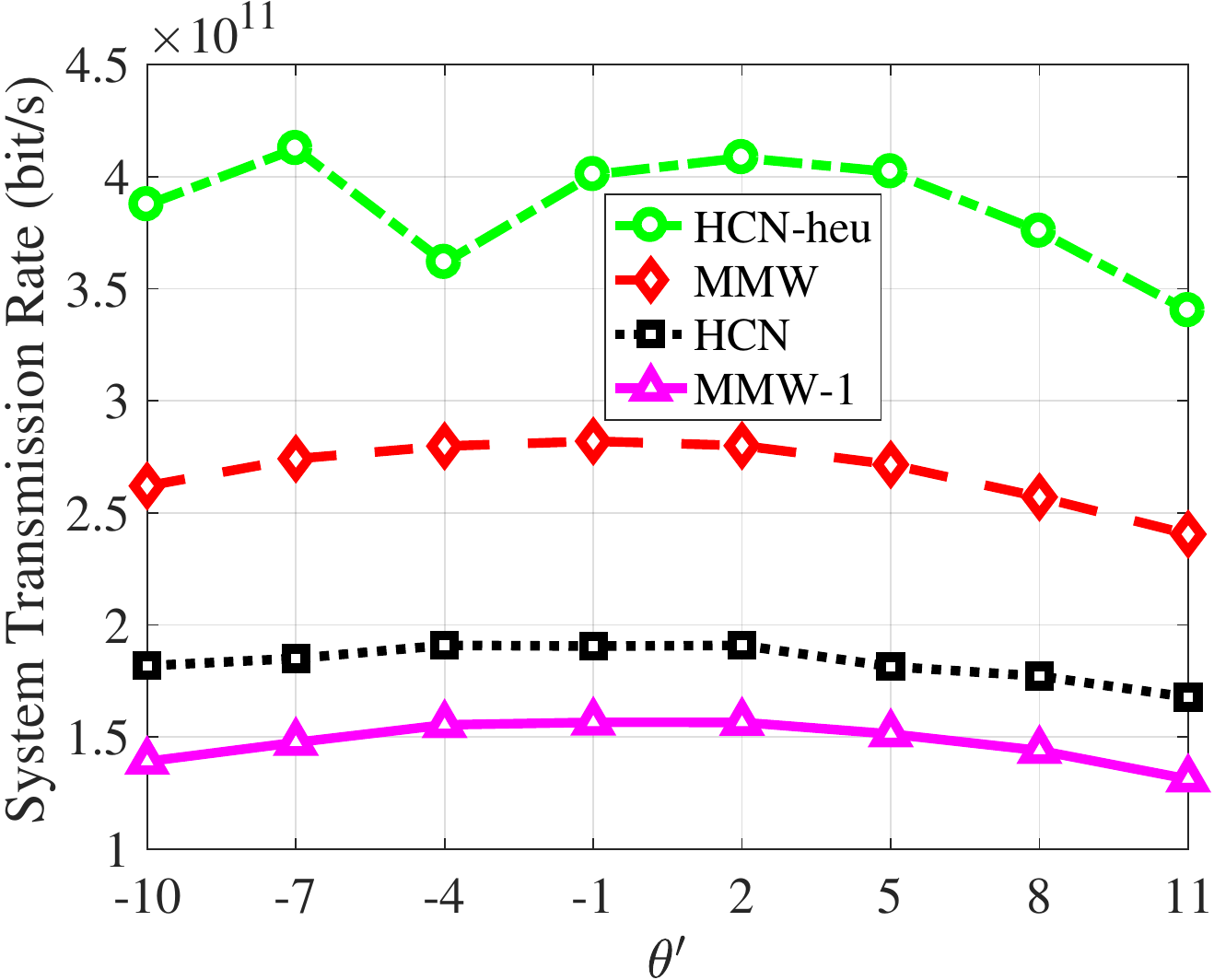}
\end{center} \caption{System transmission rate comparison of four resource allocation algorithms with different $\theta^{\prime}$.}
\label{fig10}
\end{figure}

$4) \ Mm-wave \ transmission \ power \ P_{m}:$ In Fig. \ref{fig6}, the number of cellular bands and mm-wave bands are all set to be 3. Then, we plot the system transmission rate of four schemes with the mm-wave transmission power $P_{m}$ varied from 0 to 30 dBm. As we can observe, our scheme achieves the highest system rate among the four schemes. With $P_{m}$ increased, the useful power of the numerator part and the interference power of the denominator part are also increasing, so the four curves rise slowly. Especially we continue to increase $P_{m}$ when it is already large, the effect is no longer obvious. For the proposed heuristic algorithm, the gap at $P_{m}$ of 30 dBm is only about $30\%$ of the MMW and about $105\%$ of the HCN, respectively.

$5) \ Blockage \ Probability:$  In Fig. \ref{fig7}, we plot the system transmission rate comparison of four algorithms varying $\beta$ from 0.02 to 0.16. Not surprisingly, the proposed scheme still keeps at a high level in the total transmission rate. The parameter of blockage probability is used to represent the density and size of obstacles. Therefore, when $\beta$ increases, the impact of obstacles on the mm-wave links will be greater, resulting in a decrease in the total system rate of the four algorithms involving mm-wave D2D networks. At $\beta$ of 0.16, the gap between HCN-heu and MMW is about $51\%$ of MMW, and between HCN-heu and HCN is about $137\%$ of HCN.

$6) \ {\theta}_{-3dB}:$ In Fig. \ref{fig8}, we plot the system transmission rate comparison of four algorithms by varying ${\theta}_{-3dB}$ from 10 to 80. Compared with other three algorithms, our scheme still achieves comparable performance. In the widely used realistic mm-wave directional antenna model in our paper, the parameter ${\theta}_{-3dB}$ denotes the angle of the half-power beamwidth, and there is a linear relationship between it and the main lobe width. The larger the value of ${\theta}_{-3dB}$, the greater the range that may cause interferences. Thus, with ${\theta}_{-3dB}$ increasing, the algorithms HCN-heu, MMW, HCN and MMW-1 will show a downward trend. When the ${\theta}_{-3dB}$ is equal to 80, the total transmission rate of HCN-heu is larger than that of MMW and HCN about $68\%$ and $110\%$, respectively. All in all, the choice of the ${\theta}_{-3dB}$ has an important impact on the system rate.

$7) \ Small \ Cell \ Radius:$ In Fig. \ref{fig9}, we plot the system transmission rate comparison of four algorithms by varying $r$ from 5 to 40. When the small cell radius $r$ increases, the distance between the cellular user and its serving BS will also be larger in most cases. In addition, the distance between users of the same cell and different cells may also increase, which makes the path-loss associated with distance increased for heterogeneous networks and mm-wave networks. On the other hand, the greater small cell radius $r$ will increase the blockage probability of the mm-wave links. All of these reasons make the four algorithms decreased with $r$ increasing. When $r$ is equal to 40, the total transmission rate of HCN-heu is larger than that of MMW and HCN about $43\%$ and $108\%$, respectively.

\begin{figure*}[t]
\begin{minipage}[t]{0.5\linewidth}
\centering
\includegraphics[width=0.8\columnwidth,height=2.3in]{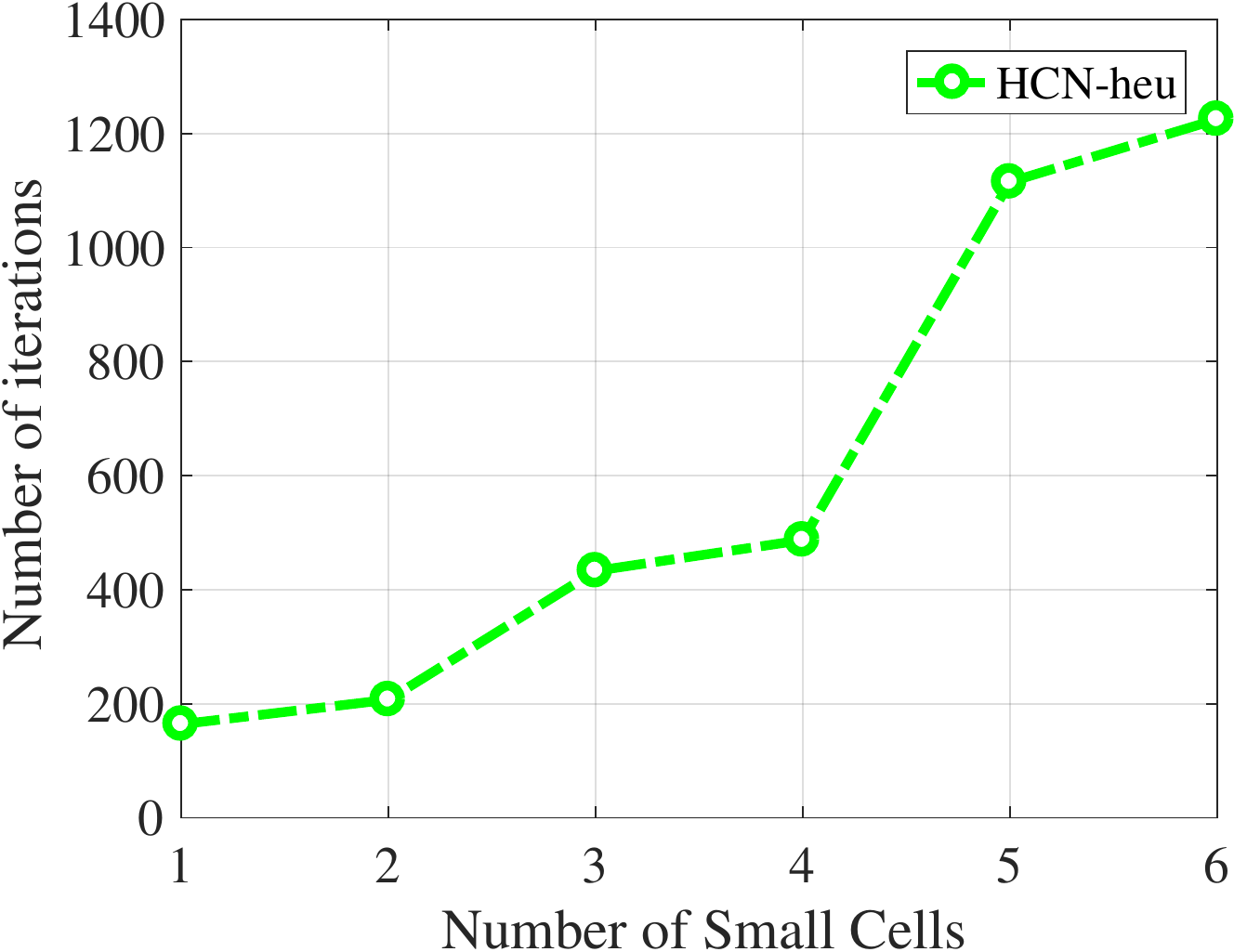}
\centerline{\small (a)}
\end{minipage}%
\begin{minipage}[t]{0.5\linewidth}
\centering
\includegraphics[width=0.8\columnwidth,height=2.3in]{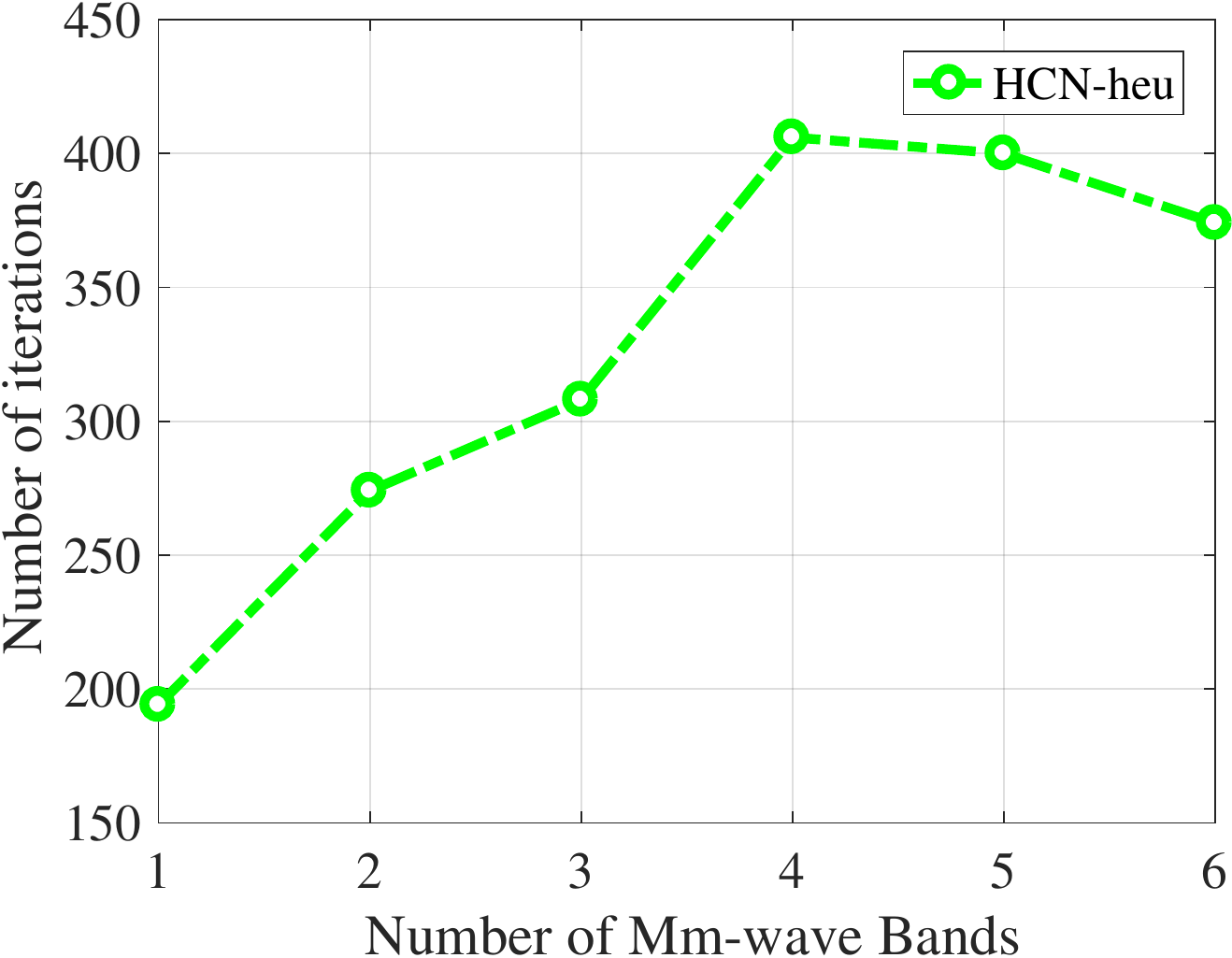}
\centerline{\small (b)}
\end{minipage}
\caption{Convergence rate in terms of number of iterations with (a) different numbers of small cells and (b) different numbers of mm-wave bands.}
\vspace*{-3mm}
\label{fig11}
\end{figure*}

$8) \ Beam \ misalignment \ error:$ All above simulations evaluate the system performance in the case of mm-wave beam alignment. Considering the important issue that the orientations of directional antennas may have a significant impact on the interferences, in Fig. \ref{fig10}, we simulate the impact of the angle of mm-wave beamforming misalignment error, which is denoted as ${\theta}^{\prime}$. The orientation of ${\theta}^{\prime}$ needs to be considered, and we assume that the clockwise orientation is positive and the counterclockwise orientation is negative. From the four algorithms, we can observe that when ${\theta}^{\prime}$ keeps approaching $0^{\circ}$ from the negative direction, the system transmission rate increases, while the system performance decreases when ${\theta}^{\prime}$ is continuously away from $0^{\circ}$ in the positive direction. Whether it is clockwise or counterclockwise, the closer the ${\theta}^{\prime}$ is to $0^{\circ}$, the larger the useful power gain and the interference power gain received by the link. In other words, the numerator and denominator of SINR are correspondingly increased, but the dominant factor is the change of numerator. Thus, the link rate will increase and the system performance will improve. Why individual points do not satisfy the above rules, which may be related to the random selection of user positions.

\subsection{Convergence Rate}\label{S5-4}

In this subsection, we analyze the convergence rate by simulating the number of iterations. Both the number of cellular bands and the number of mm-wave bands are set to be 3, the maximum number of D2D pairs of each cell to be 15, and we vary the number of cells from 1 to 6 to obtain the simulation results in Fig. \ref{fig11}(a). From the figure, we observe that the number of iterations increases with the number of cells. When the horizontal axis takes 4, comparing with the iterations of the exhaustive search method from $6^{4\times 1}$ to $6^{4\times 15}$, the iterations of the proposed scheme is 487 and the complexity is greatly reduced. The same conclusion applies to other values. In addition, we set the number of cellular bands and small cells to be 2 and 3, respectively. The simulation results are shown in Fig. \ref{fig11}(b). When the number of mm-wave bands is sufficiently large, the number of D2D pairs in each band set is small when initial resource allocation is performed, so less iterations is required to reach a near-optimal solution.

\section{Conclusion}\label{S6}

In this paper, we have investigated the uplink resource allocation problem for D2D communications underlaying the scenario of multi-band heterogeneous cellular networks with small cells densely deployed. Considering complex intra- and inter-cell interferences, we formulate the optimization problem of resource allocation among multiple micro-wave bands and mm-wave bands for multiple D2D pairs. The optimization metric is maximizing the system performance in terms of system transmission rate. To cope with this issue, we propose a heuristic algorithm to yield the near-optimal solution with fairly low computational complexity. Finally, through extensive simulations under various system parameters, and analysis of the optimality and complexity of our proposed algorithm, we complete the demonstration of the superior performance of the proposed scheme. Averagely, the proposed algorithm improves the system performance by about $36\%$ compared with full mm-wave transmission scheme. In the future work, we will take the user mobility into account to make our study more practical.

\bibliographystyle{IEEEtran}

\begin{thebibliography}{10}

\bibitem{Qualcomm}
\emph{Qualcomm Data Challenge}, accessed: Sep. 2013. [Online].
Available: https://www.qualcomm.com/documents/rising-meet-1000x-mobile-data-challenge

\bibitem{niu2017energy}
Y.~Niu, C.~Gao, Y.~Li, L.~Su, D.~Jin, Y.~Zhu, and D.~O. Wu, ``Energy-efficient
  scheduling for mmwave backhauling of small cells in heterogeneous cellular
  networks,'' \emph{IEEE Transactions on Vehicular Technology}, vol.~66, no.~3,
  pp. 2674--2687, Mar. 2017.

\bibitem{Future}
B.~Ai, K.~Guan, M.~Rupp, T.~K\"urner, X.~Cheng, X.~F. Yin, Q.~Wang, G.~Y. Ma, Y.~Li, L.~Xiong, and J.~W. Ding, ``Future Railway Services-Oritend Mobile Communications Network,''
\emph{IEEE Communications Magazine}, vol.~53, no.~10, pp. 78--85, Oct. 2015.

\bibitem{Ruisi}
R.~S. He, B.~Ai, G.~L. Stuber, G.~P. Wang, and Z.~D. Zhong, ``Geometrical based modeling for millimeter wave MIMO mobile-to-mobile channels,'' \emph{IEEE Transactions on Vehicular Technology}, vol.~67, no.~4, pp. 2848--2863, Nov. 2017.

\bibitem{ECMA387}
\emph{High Rate 60 GHz PHY, MAC and HDMI PAL}, ECMC TC48, ECMA Std. 387, Dec. 2008.

\bibitem{IEEE802153c}
\emph{IEEE 802.15 WPAN Millimeter Wave Alternative PHY Task Group 3c
(TG3c)}. Accessed: May 20, 2017. [Online]. Available: http://www.ieee802.org/15/pub/TG3c.html

\bibitem{spaceair}
J.~J. Liu, Y.~P. Shi, Z.~Fadlullah, and N.~Kato, ``Space-Air-Ground Integrated Network: A Survey,'' \emph{IEEE Communications Surveys \& Tutorials}, vol. Early Access, 2018.

\bibitem{niu2018}
Y.~Niu, Y.~Liu, Y.~Li, X.~L. Chen, Z.~D. Zhong, and Z.~Han, ``Device-to-Device Communications Enabled Energy Efficient Multicast Scheduling in mmWave Small Cells,'' \emph{IEEE Transactions on Communications}, vol.~66, no.~3, pp. 1093--1109, Mar. 2018.

\bibitem{FiWi}
J.~J. Liu, H.~Z. Guo, H.~Nishiyama, H.~Ujikawa, K.~Suzuki, and N.~Kato, ``New Perspectives on Future Smart FiWi Networks: Scalability, Reliability, and Energy Efficiency,'' \emph{IEEE Communications Surveys \& Tutorials}, vol.~18, no.~2, pp. 1045--1072, Nov. 2015.

\bibitem{mmWavesub6}
J.~Q. Deng, O.~Tirkkonen, R.~F. Hollanti, T.~Chen, and N.~Nikaein, ``Resource Allocation and Interference Management for Opportunistic Relaying in Integrated mmWave/sub-6 GHz 5G Networks,'' \emph{IEEE Communications Magazine},
  vol.~55, no.~6, pp. 94--101, Jun. 2017.

\bibitem{ai2017indoor}
B.~Ai, K.~Guan, R.~S. He, J.~Z. Li, G.~K. Li, D.~P. He, Z.~D. Zhong, and K.~Huq, ``On indoor millimeter wave massive mimo channels: Measurement and simulation,'' \emph{IEEE Journal on Selected Areas in Communications},
  vol.~35, no.~7, pp. 1678--1690, Jul. 2017.

\bibitem{YongNiu}
Y.~Niu, C.~Gao, Y.~Li, L.~Su, D.~Jin, and A.~V. Vasilakos, ``Exploiting Device-to-Device Communications in Joint Scheduling of Access and Backhaul for mmWave Small Cells,'' \emph{IEEE Journal on Selected Areas in Communications}, vol.~33, no.~10, pp. 2052--2069, May. 2015.

\bibitem{singh2011interference}
S.~Singh, R.~Mudumbai, and U.~Madhow, ``Interference analysis for highly
  directional 60-ghz mesh networks: The case for rethinking medium access
  control,'' \emph{IEEE/ACM Transactions on Networking (TON)}, vol.~19, no.~5,
  pp. 1513--1527, Oct. 2011.

\bibitem{jiajia1}
J.~J. Liu, Y.~C. Kawamoto, H.~Nishiyama, N.~Kato, and N.~Kadowaki, ``Device-to-device communications achieve efficient load balancing in LTE-advanced networks,'' \emph{IEEE Wireless Communications}, vol.~21, no.~2, pp. 57--65, Apr. 2014.

\bibitem{jiajia2}
J.~J. Liu, N.~Kato, J.~F. Ma, and N.~Kadowaki, ``Device-to-Device Communication in LTE-Advanced Networks: A Survey,'' \emph{IEEE Communications Surveys \& Tutorials}, vol.~17, no.~4, pp. 1923--1940, Dec. 2014.

\bibitem{jiajia3}
J.~J. Liu, S.~W. Zhang, N.~Kato, H.~Ujikawa, and K.~Suzuki, ``Device-to-device communications for enhancing quality of experience in software defined multi-tier LTE-A networks,'' \emph{IEEE Network}, vol.~29, no.~4, pp. 46--52, Jul. 2015.


\bibitem{ma2015interference}
C.~Ma, J.~Liu, X.~Tian, H.~Yu, Y.~Cui, and X.~Wang, ``Interference exploitation
  in d2d-enabled cellular networks: A secrecy perspective,'' \emph{IEEE
  Transactions on Communications}, vol.~63, no.~1, pp. 229--242, Jan. 2015.

\bibitem{ramezani2017joint}
A.~R. Kebrya, M.~Dong, B.~Liang, G.~Boudreau, and S.~H. Seyedmehdi,
  ``Joint power optimization for device-to-device communication in cellular
  networks with interference control,'' \emph{IEEE Transactions on Wireless
  Communications}, vol.~16, no.~8, pp. 5131--5146, Aug. 2017.


\bibitem{li2014coalitional}
Y.~Li, D.~Jin, J.~Yuan, and Z.~Han, ``Coalitional games for resource allocation
  in the device-to-device uplink underlaying cellular networks,'' \emph{IEEE
  Transactions on Wireless Communications}, vol.~13, no.~7, pp. 3965--3977,
  Jul. 2014.

\bibitem{xu2013efficiency}
C.~Xu, L.~Song, Z.~Han, Q.~Zhao, X.~Wang, X.~Cheng, and B.~Jiao, ``Efficiency
  resource allocation for device-to-device underlay communication systems: A
  reverse iterative combinatorial auction based approach,'' \emph{IEEE Journal
  on Selected Areas in Communications}, vol.~31, no.~9, pp. 348--358, Sep. 2013.

\bibitem{Dai}
J.~H. Dai, J.~J. Liu, Y.~P. Shi, S.~B. Zhang, and J.~F. Ma, ``Analytical Modeling of Resource Allocation in D2D Overlaying Multihop Multichannel Uplink Cellular Networks,'' \emph{IEEE Transactions on Vehicular Technology}, vol.~66, no.~8, pp. 6633--6644, Aug. 2017.

\bibitem{InterferenceCoordination}
H.~Wang, S.~H. Leung, and R.~F. Song, ``Uplink Area Spectral Efficiency Analysis for Multichannel Heterogeneous Cellular Networks With Interference Coordination,'' \emph{IEEE Access}, vol.~6, pp. 14485--14497, Mar. 2018.

\bibitem{jointaccess}
Z.~Y. Tan, X.~Li, F.~R. Yu, L.~Chen, H.~Ji, and V.~Leung, ``Joint Access Selection and Resource Allocation in Cache-Enabled HCNs with D2D Communications,'' in \emph{IEEE Wireless Communications and Networking Conference (WCNC)}, San Francisco, CA, USA, Mar. 2017, pp. 1--6.

\bibitem{zhenxiang}
Z.~X. Su, B.~Ai, D.~P. He, G.~Y. Ma, K.~Guan, N.~Wang, and D.~Y. Zhang, ``User Association and Backhaul Bandwidth Allocation for 5G Heterogeneous Networks in the Millimeter-Wave Band,'' in \emph{IEEE/CIC International Conference on Communications in China (ICCC)}, Qingdao, China, Oct. 2017, pp. 1--6.

\bibitem{niu2015}
Y.~Niu, C.~H. Gao, Y.~Li, L.~Su, D.~P. Jin, and A.~V. Vasilakos, ``Exploiting Device-to-Device Communications in Joint Scheduling of Access and Backhaul for mmWave Small Cells,'' \emph{IEEE Journal on Selected Areas in Communications}, vol.~33, no.~10, pp. 2052--2069, May. 2015.

\bibitem{niuyy}
Y.~Niu, H.~Shi, Y.~Li, R.~S. He, and Z.~D. Zhong ``Coalition Formation Games Based Sub-Channel Allocation for Device-to-Device Underlay mmWave Small Cells,'' in \emph{2nd URSI Atlantic Radio Science Meeting (AT-RASC)}, Meloneras, Spain, May. 2018.

\bibitem{yali}
Y.~L. Chen, B.~Ai, Y.~Niu, K.~Guan, and Z.~Han, ``Resource Allocation for Device-to-Device Communications Underlaying Heterogeneous Cellular Networks Using Coalitional Games,'' \emph{IEEE Transactions on Wireless Communications}, vol.~17, no.~6, pp. 4163--4176, Jun. 2018.

\bibitem{Zhenyu}
Z.~Y. Xiao, P.~F. Xia, and X.~G. Xia, ``Codebook Design for Millimeter-Wave Channel Estimation with Hybrid Precoding Structure,'' \emph{IEEE Transactions on Wireless Communications}, vol.~16, no.~1, pp. 141--153, Jan. 2017.

\bibitem{Zhenyu2}
Z.~Y. Xiao, P.~F. Xia, and X.~G. Xia, ``Full-Duplex Millimeter-Wave Communication,'' \emph{IEEE Wireless Communications}, vol.~24, no.~6, pp. 136--143, Dec. 2017.

\bibitem{Guan2017}
K.~Guan, G.~K. Li, T.~K\"urner, A.~F. Molisch, B.~Peng, R.~S. He, B.~Hui, J.~Kim, and Z.~D. Zhong, ``On Millimeter Wave and THz Mobile Radio Channel for Smart Rail Mobility,'' \emph{IEEE Transactions on Vehicular Technology}, vol.~66, no.~7, pp. 5658--5674, Jul. 2017.

\bibitem{Zhenyu3}
Z.~Y. Xiao, T.~He, P.~F. Xia, and X.~G. Xia, ``Hierarchical Codebook Design for Beamforming Training in Millimeter-Wave Communication,'' \emph{IEEE Transactions on Wireless Communications}, vol.~15, no.~5, pp. 3380--3392, May. 2016.

\bibitem{Zhenyu4}
Z.~Y. Xiao, L.~P. Zhu, J.~Choi, P.~F. Xia, and X.~G. Xia, ``Joint Power Allocation and Beamforming for Non-Orthogonal Multiple Access (NOMA) in 5G Millimeter Wave Communications,'' \emph{IEEE Transactions on Wireless Communications}, vol.~17, no.~5, pp. 2961--2974, Feb. 2018.

\bibitem{mmwchannel}
M.~R. Akdeniz, Y.~P. Liu, M.~K. Samimi, S.~Sun, S.~Rangan, T.~S. Rappaport, and E.~Erkip, ``Millimeter wave channel modeling and cellular capacity evaluation,'' \emph{IEEE Journal on Selected Areas in Communications}, vol. 32, no. 6, pp. 1164--1179, Jun. 2014.

\bibitem{BlockageRobust}
Y.~Niu, Y.~Li, D.~Jin, L.~Su, and D.~O. Wu, ``Blockage Robust and Efficient Scheduling for Directional mmWave WPANs,'' \emph{IEEE Transactions on Vehicular Technology}, vol.~64, no.~2, pp. 728--742, Feb. 2015.

\bibitem{CoverageAnalysis}
X.~H. Yu, J.~Zhang, M.~Haenggi, and K.~B. Letaief, ``Coverage Analysis for Millimeter Wave Networks: The Impact of Directional Antenna Arrays,'' \emph{IEEE Journal on Selected Areas in Communications}, vol.~35, no.~7, pp. 1498--1512, Apr. 2017.

\bibitem{channelmodel}
C.~Ma, M.~Ding, D.~L. P\'erez, Z.~H. Lin, J.~Li, and G.~Q. Mao, ``Performance Analysis of the Idle Mode Capability in a Dense Heterogeneous Cellular Network,'' \emph{IEEE Transactions on Communications}, vol. Early Access, 2018.

\bibitem{probability}
J.~J. Liu, H.~Nishiyama, N.~Kato, and J.~Guo, ``On the Outage Probability of Device-to-Device-Communication-Enabled Multichannel Cellular Networks: An RSS-Threshold-Based Perspective,'' \emph{IEEE Journal on Selected Areas in Communications}, vol.~34, no.~1, pp. 163--175, Jan. 2016.

\bibitem{REX}
L.~X. Cai, L.~Cai, X.~M. Shen, and J.~W. Mark, ``Rex: A randomized Exclusive region based scheduling scheme for mmWave WPANs with directional antenna,'' \emph{IEEE Transactions on Wireless Communications}, vol.~9, no.~1, pp. 113--121, Jan. 2010.


\bibitem{Lee2016Connectivity}
I.~H. Lee and H.~Jung, ``Connectivity analysis of millimeter-wave
  device-to-device networks with blockage,'' \emph{International Journal of
  Antennas and Propagation}, vol.~2016, Oct. 2016, Art. no.~7939671.

\bibitem{pisinger2005hard}
D.~Pisinger, ``Where are the hard knapsack problems?'' \emph{Computers \&
  Operations Research}, vol.~32, no.~9, pp. 2271--2284, 2005.

\bibitem{antennamodel}
Q.~Chen, X.~Peng, J.~Yang, and F.~Chin, ``Spatial reuse strategy in mmWave WPANs with directional antennas,'' in \emph{IEEE Global Communications Conference (GLOBECOM)}, Anaheim, CA, Dec. 2012, pp. 5392--5397.

\end{thebibliography}

\end{document}